\documentclass[fleqn,usenatbib]{mnras}

\usepackage[T1]{fontenc}

\usepackage{graphicx}	
\usepackage{amsmath}	
\usepackage{amssymb}	
\usepackage{subcaption}
\usepackage{cleveref}
\crefname{figure}{Fig.}{Figs.}
\crefname{table}{Table}{Tables}
\crefname{chapter}{Chapter}{Chapters}
\crefname{section}{Section}{Sections}
\usepackage{newtxtext,newtxmath}

\newcommand{\MSUN}{{\rm M}_\odot}
\newcommand{\MSTAR}{M_{\rm *}}

\newcommand{\MHOST}{M_{\rm 200c}}
\newcommand{\MDYN}{M_{\rm dyn}}
\newcommand{\Mallgravgas}{M^\mathrm{allgrav}_\mathrm{gas}}
\newcommand{\Mallgravstar}{M^\mathrm{allgrav}_*}
\newcommand{\rhStar}{r_{1/2,\ast}}
\usepackage[dvipsnames]{xcolor}

\title[TNG50 jellyfish galaxies and their star formation]{Jellyfish galaxies with the IllustrisTNG simulations -- No enhanced population-wide star formation according to TNG50}

\author[J. Göller et al.]{
Junia Göller$^{1,2}$\thanks{junia.goeller@uni-heidelberg.de}, Gandhali D. Joshi$^{2,3}$, Eric Rohr$^{2}$, Elad Zinger$^{2,4}$, and Annalisa Pillepich$^{2}$
\\
$^{1}$Institut f\"ur Theoretische Astrophysik, Universit\"at Heidelberg, Albert-Ueberle-Straße 2, D-69120 Heidelberg, Germany\\
$^{2}$Max-Planck-Institut f\"ur Astronomie, K\"onigstuhl 17, D-69117 Heidelberg, Germany\\
$^{3}$Department of Physics and Astronomy, University College London, London WC1E 6BT, UK\\
$^{4}$Centre for Astrophysics and Planetary Science, Racah Institute of Physics, The Hebrew University, Jerusalem 91904, Israel
}

\date{Accepted XXX. Received YYY; in original from ZZZ}

\pubyear{2023}

\begin{document}
\label{firstpage}
\pagerange{\pageref{firstpage}--\pageref{lastpage}}
\maketitle

\begin{abstract}
Due to ram-pressure stripping, jellyfish galaxies are thought to lose large amounts, if not all, of their interstellar medium. Nevertheless, some, but not all, observations suggest that jellyfish galaxies exhibit enhanced star formation compared to control samples, even in their ram pressure-stripped tails. We use the TNG50 cosmological gravity+magnetohydrodynamical simulation, with an average spatial resolution of 50-200 pc in the star-forming regions of galaxies, to quantify the star formation activity and rates (SFRs) of more than 700 jellyfish galaxies at $z=0-1$ with stellar masses $10^{8.3-10.8}\,\MSUN$ in hosts with mass $10^{10.5-14.3}\,\MSUN$. We extract their global SFRs, the SFRs within their main stellar body vs. within the tails, and we follow the evolution of the star formation along their individual evolutionary tracks. We compare the findings for jellyfish galaxies to those of diversely-constructed control samples, including against satellite and field galaxies with matched redshift, stellar mass, gas fraction and host halo mass. According to TNG50, star formation and ram-pressure stripping can indeed occur simultaneously within any given galaxy, and frequently do so. Moreover, star formation can also take place within the ram pressure-stripped tails, even though the latter is typically subdominant. However, TNG50 does not predict elevated population-wide SFRs in jellyfish compared to analogue satellite galaxies with the same stellar mass or gas fraction. Simulated jellyfish galaxies do undergo bursts of elevated star formation along their history but, at least according to TNG50, these do not translate into a population-wide enhancement at any given epoch.
\end{abstract}




\section{Introduction}
\label{sec:intro}
Over the last few decades, observations have shown that galaxies in high-density environments, such as satellites within massive groups and clusters, have different properties than galaxies in the field, i.e. in lower-density environments. For example, group and cluster galaxies are H$\,$I deficient \citep{GiovanelliHaynes1985}, have redder colours \citep{Kennicutt1983} and a lower star formation rate \citep[SFR,][]{BowerBalogh2004}, and exhibit non-disky stellar morphologies more frequently \citep{Dressler1980} than similar-mass analogues in the field.

Several different mechanisms have been proposed to explain these observational findings, with the key, common idea being that the interaction between satellites and their environment reduces their gas availability, including that of their interstellar medium (ISM). For example, it is commonly thought that for group and cluster galaxies, the replenishment of gas for star formation is largely suppressed \citep{BoselliFossati2022}. This can occur either because gas inflows are impaired by the satellites' large orbital velocities through the intra-group/cluster medium (IGM/ICM) or because the satellites' outer gas reservoirs (their circum-galactic media, CGM) are stripped \citep[starvation or strangulation,][]{Larson1980}. Ram-pressure stripping \citep[RPS,][]{Gunn1972} has been shown to be capable of pulling out of satellite galaxies not only the loosely-bound CGM, but also the cold and dense ISM, producing tails of gas in the direction opposite to a galaxy's direction of motion. 

Galaxies in the process of being ram-pressure stripped and exhibiting strongly asymmetrical gas distributions have been dubbed {\it jellyfish galaxies} in the literature \citep{Ebeling2014a}, with wakes of gas departing from their central luminous bodies. Ram-pressure stripped and jellyfish galaxies have been observed from the X-ray \citep{Sun2010} to the radio \citep{Roberts2021}. A recent and systematic effort dedicated to the investigation of jellyfish galaxies is the GASP survey \citep[GAs Stripping Phenomena in galaxies with MUSE;][]{Poggianti2017}, which includes 114 ram-pressure stripped galaxies at redshift $z=0.04-0.07$. Jellyfish galaxies have also been studied with the VIVA \citep[VLA Imaging of Virgo spirals in Atomic gas;][]{Chung2009}, MACS \citep[MAssive Clsuter Survey;][]{Ebeling2014a}, and LoTSS \citep[LOFAR Two-Meter Sky Survey;][]{Roberts2021} surveys. Moreover, several individual galaxies have been examined at various wavelengths in great detail, e.g. ESO$\,$137-001 in the Norma cluster \citep{Sun2010, Fumagalli2014, Jachym2019} or D100 in the Coma cluster \citep{Yagi2007, Cramer2019}.

From observations, it is apparent that jellyfish galaxies are not fully quenched but are instead star forming, even though the ongoing RPS reduces their gas and ISM content. Star formation may happen in the compressed gas in the body of the galaxies \citep{Vulcani2018, Roberts22}, as well as in their tails \citep{Vulcani2018, Cramer2019, Jachym2019}. In the tails, star formation sometimes appears in so called 'fireballs' \citep{Jachym2019}, i.e. star forming regions producing streams of young stars that extend in the direction of the galactic body. How and why star formation is observed in the diffuse gas of the galactic tails -- which have much lower densities than the disks -- remains an open question. Magnetic fields are often predicted to play a role, particularly in enabling star formation in the underdense gas of jellyfish tails \citep{Safarzadeh2019, Mueller2021}. Yet, causal connections are difficult to pinpoint, because star formation is not a linear process and complex internal dynamics have to be taken into account in the case of jellyfish galaxies \citep{Roediger2009a}.

Some observations have even suggested globally-enhanced SFRs in jellyfish galaxies \citep{Vulcani2018, Ramatsoku2020, Vulcani2020} in comparison to other samples of satellite or field galaxies. This possibility would further reinforce the role of jellyfish galaxies as a connection between field and cluster galaxies, as some features in the typical Balmer lines of cluster satellites can only be explained if their quenching is preceded by a burst in star formation \citep{Abadi1999}. However, other observations have shown no signs of enhanced SFR or have even found it reduced in jellyfish galaxies \citep{Yoon2017, Mun2021}. The question of whether jellyfish have enhanced SFRs compared to satellite or field analogues is still open. 

In fact, whereas most of those analyses use H$\alpha$ as a tracer of ongoing star formation, it has been suggested that H$\alpha$ in RPS tails may not be as tightly linked to star formation as in galactic disks \citep{Boselli2016}, with the excitation instead being caused by shocks or heat conduction. For this reason, measurements of star formation from H$\alpha$ intensities could lead to an overestimation of the SFR in the tails \citep{Cramer2019}.
Therefore, reliable measurements of the star formation activity in the bodies and tails of observed jellyfish galaxies remain missing and the SFRs and SF efficiencies of jellyfish compared to non-stripped galaxies are still debated. 

Given the difficulties in extracting SFRs observationally and the limited number of observed jellyfish galaxies and their complex selection functions, theoretical insights can provide guidance. Many efforts have been made over the last decade to simulate jellyfish galaxies, with a variety of codes, techniques, and included physical processes \citep[e.g.][]{Kronberger2008, Kapferer2009, Tonnesen2012, Roediger2014, Steinhauser2016, TroncosoIribarren2016, Ramos-Martinez2018, Yun2019a, Troncoso-Iribarren2020}. 

In relation to the critical process of star formation in jellyfish galaxies, and to star formation within bodies (i.e. the main stellar component and the corresponding gas reservoir) and tails, different simulation approaches have given different outcomes. RPS enhances the overall SFR  in the work by \citealt{Kronberger2008} and \citealt{Kapferer2009}, who simulated the interaction of an individual disk galaxy with its environment with the smooth-particle-hydordynamics (SPH) code \textsc{Gadget-2}: for \citealt{Kronberger2008}, including models for cooling, star formation, stellar feedback and galactic winds, stars mainly form in the disks within the compressed central regions and in the stripped material behind the disk, whereas \citealt{Kapferer2009}, accounting for cooling, star formation and stellar feedback in galaxies of various different infall velocities in a surrounding medium of differing density, found a suppression of SFR in the disk and a shift of SFR from disk to tail. For \citealt{Tonnesen2012}, by using the adaptive mesh refinement code \textsc{Enzo} to model a massive spiral galaxy in a constant, face-on ICM flow while accounting for radiative cooling, star formation and Type II supernova feedback, star formation is truncated in the galactic disk, slightly enhanced in the bulge, and only low levels of star formation are found in the wake of the galaxy. \citealt{Roediger2014}, in an idealized hydrodynamical simulation with the grid code \textsc{Flash}, including radiative cooling, star formation and feedback in a disk galaxy that is exposed to a constant ICM flow, predicted an enhancement of SFR in disk regions where the gas will be stripped in the near future, i.e. in lower-density regions, and star formation in gas knots in the tail. By simulating a galaxy cluster with the moving-mesh code \textsc{Arepo}, modeling cooling, star formation and feedback, \citealt{Steinhauser2016} found a general enhancement of SFR only for galaxies experiencing mild ram pressure. \citealt{Ramos-Martinez2018}, in a magnetohydrodynamical simulation of a magnetized disk galaxy experiencing face-on RPS using the adaptive mesh refinement code \textsc{Ramses}, showed that magnetic fields can channel gas to the centre of the galaxy where it can be a reservoir for star formation.

These simulations \citep[with the exception of][]{Steinhauser2016} all relied on wind-tunnel setups, where RPS is studied in individual galaxies with great control on the initial conditions and with the possibility of following the effects of many detailed physical processes within the galaxies at high numerical resolution. The influence of galaxy type, angle of infall, and inclusion of gas cooling and magnetic fields, are just a few examples of the types of examinations carried out in these numerical experiments. However, they come at the cost of the broader, larger-scale picture, as they lack physical processes such as galaxy-galaxy interactions and pre-processing, and do not reproduce the large diversity and number statistics that characterize galaxy populations in the Universe. 

Full cosmological galaxy simulations such as EAGLE \citep{Crain2015, Schaye2015} and IllustrisTNG \citep{Springel2018, Naiman2017, Marinacci2017, Pillepich2017, Pillepich2018, Pillepich2019, Nelson2018, Nelson2019}, instead, follow the formation and evolution of tens of thousands of galaxies from the initial conditions of the Universe shortly after the Big Bang, and naturally take into account the hierarchical growth of structure and the mutual interactions among galaxies and of galaxies with the larger-scale structure. In these simulations, cosmological gas accretion, galaxy-galaxy mergers and interactions, tidal and RPS, gravitational heating, etc. all emerge naturally from the numerical solution of the coupled equations of gravity and hydrodynamics in expanding synthetic universes. These simulations therefore  form and evolve jellyfish galaxies in a self-consistent way when satellites orbit and interact with the dense intra-halo gas within groups and clusters of galaxies \citep[][for IllustrisTNG and EAGLE, respectively]{Yun2019a, Troncoso-Iribarren2020}. For example, \citealt{TroncosoIribarren2016} and \citealt{Troncoso-Iribarren2020} have shown that the EAGLE simulation predicts an enhancement of star formation in the so-called leading half of a galaxy falling into a cluster, whereas in the trailing half no increase in SFR is found.  

In this paper, we analyse the TNG50 simulation \citep{Pillepich2019, Nelson2019}, the highest-resolution flagship run of IllustrisTNG, to investigate the connection between jellyfish status and star formation activity within the boundary conditions of the IllustrisTNG model: this includes, among others, gas cooling and heating, feedback from stars and super massive black holes, and effective recipes for the conversion of gas into stars and for the thermodynamics of the ISM. TNG50 reaches average spatial gas resolutions of 50-200 pc in the star-forming regions of galaxies \citep{Pillepich2019} while simultaneously following the co-evolution of thousands of galaxies, including those in the high-density environments of $10^{13-14.3}\,\MSUN$ halos. Hence, it provides a large sample of galaxies, across diverse evolutionary stages, environments, and properties, which in turn allow for the statistically-robust characterization of the star formation activity of about 700 jellyfish galaxies at $z=0-1$ in comparison to variously-constructed control samples.

This paper is a companion paper of \citealt{JelZinger} and \citealt{JelRohr}. In the former, we describe the process of jellyfish galaxy identification in the TNG50 and TNG100 simulations (TNG100 being another flagship run of the IllustrisTNG series), which have been carried out visually via the ``Cosmological Jellyfish'' Zooniverse project: we compare the identification outcomes of this citizen science project with that of professionals, as well as provide a first analysis of the jellyfish galaxies demographics. \citealt{JelRohr} analyses the evolutionary properties of jellyfish galaxies and investigates the cold gas loss during their life cycle.

This paper is organized as follows: in Section~\ref{sec:tng50} we outline the numerical model used in the TNG50 simulation. How we identify jellyfish galaxies is described in Section~\ref{sec:zooniverse}. In Section~\ref{sec:galaxymasses}, \ref{sec:sfr} and \ref{sec:samples} we define how we measure galaxy stellar mass, SFR and the different galaxy samples used throughout, whereas in \ref{sec:tracking} we explain how we track individual galaxies throughout cosmic epochs. We then study the general demographics of our galaxy samples in Section~\ref{sec:demographics} before we do a detailed examination of the galaxies' SFRs in Section~\ref{sec:disk_tail} and \ref{sec:global_sfr}. Finally we investigate the evolution of SFR over time within bins of galactic mass (Section~\ref{sec:SFRredshift}) and in individual galaxies (Section~\ref{sec:SFRbursts}). We discuss our results and simulation details in Section~\ref{sec:discussion} and conclude and summarize in Section~\ref{sec:summary}.

\section{Methods and data}

\subsection{TNG50 and the IllustrisTNG numerical model}
\label{sec:tng50}
TNG50 \citep{Pillepich2019, Nelson2019} is a cosmological gravity+magnetohydrodynamics simulation for the formation and evolution of galaxies in the $\Lambda$CDM cosmological scenario. It is the highest-resolution run of the IllustrisTNG project\footnote{\url{www.tng-project.org}} \citep{Springel2018, Naiman2017, Marinacci2017, Pillepich2018, Nelson2018}, encompasses a periodic boundary-condition cube of (51.7$\,$ comoving Mpc)$^3$, models about 6,500 galaxies at $z=0$ with stellar mass $\gtrsim10^8\,\MSUN$ while having a target baryonic mass resolution of $8.5\times 10^4\,\MSUN$. With such properties, it lies in between the classical zoom-in and full-box regimes, making it possible to simultaneously describe large-scale phenomena such as the mutual interaction of galaxies and of galaxies within massive groups and clusters as well as  small-scale processes like star formation, from $z=127$ to the current epoch. 

In fact, TNG50 takes into account gravitational interactions on all cosmic scales down to a fraction of the softening length(s) \citep{Nelson2019} as well as the hydrodynamics of the cosmic fluid and the evolution of magnetic fields, with the moving-mesh code \textsc{Arepo} \citep{Springel2010}.

A large set of astrophysical processes are included in the simulation to follow galaxy formation and evolution, such as star formation, stellar evolution, non-local feedback from Type II supernovae, seeding and growth of super massive black holes and their feedback, chemical enrichment of the ISM, and the heating and cooling of gas, also in connection to the changing gas metallicity \citep[see][for more details on the IllustrisTNG model]{Pillepich2018, Weinberger2017}. For example, the abundance of nine chemical elements (H, He, C, N, O, Ne, Mg, Si, Fe) is tracked during the cosmic evolution, in addition to Europium, and stars are represented as mono-age stellar populations by stellar particles. Star formation and feedback are subgrid models. The size of the gas cells is adaptive, with smaller cells at higher densities, down to cells as small as 5-10 pc in the highest density regions within galaxies \citep[see][for more details]{Pillepich2019, Pillepich2021}. 

Cosmic structures, i.e. haloes and subhaloes and the galaxies therein, are identified on the fly by two methods: the Friends-of-Friends algorithm \citep[FoF][]{Davis1985}, which finds host halos based on dark matter structure, and the \textsc{Subfind} \citep{Springel2000} algorithm that identifies overdensities assembling galaxies within those halos.

\subsection{Visually-identified IllustrisTNG jellyfish}
\label{sec:zooniverse}
TNG50 contains thousands of resolved galaxies and visually classifying such large numbers is a challenging task.
Hence, we make use of the results of the ``Cosmological Jellyfish'' Zooniverse project, developed by our team and built upon the citizen-science portal ``Zooniverse''\footnote{\url{https://www.zooniverse.org/}}. All details of the jellyfish-identification process are described by \citealt{JelZinger} and succinctly summarized here.

On a dedicated website\footnote{\url{https://www.zooniverse.org/projects/apillepich/cosmological-jellyfish}}, we asked volunteers, after a short training exercise, to decide if a galaxy resembles a jellyfish or not, based on an image of its gas mass surface density and stellar mass density contours, in a random projection. Volunteers were instructed to classify objects which are in close proximity to another galaxy as \emph{non-jellyfish} even if they had tails, to exclude tails formed due to tidal stripping, since we aimed for a pure (and not necessarily complete) sample. In this way, a total of 80,704 galaxies between $z=0$ and $z=2$ from TNG50 and TNG100 were classified by more than 6,000 volunteers across two phases. We expand on the selection criteria of the inspected sample in Section~\ref{sec:samples}. Each galaxy was inspected by 20 volunteers, after which the corresponding image was retired from the classification. Since the scheme is binary in its raw output, such a visual classification scheme returned for each galaxy a score between 0 (no inspecting volunteer classified it as a jellyfish galaxy) and 20 (all inspecting volunteers classified it as a jellyfish galaxy). These raw scores are subsequently weighted by the experience of the inspectors and the agreement with expert inspectors, and the total score is normalized to a value between 0 and 1. A galaxy with a score larger than a certain threshold can be considered a jellyfish: we follow the recommendations of \citealt{JelZinger} (which again aimed for a sample of high purity) as described in the next Sections.

\subsection{Galaxy stellar and total mass}
\label{sec:galaxymasses}
Throughout this paper, we define a galaxy's stellar mass as the summed mass of stellar particles that are gravitationally bound according to \textsc{Subfind} and that are within a distance of $<2\times \rhStar$ from the center of the galaxy. Here $\rhStar$ denotes the stellar half-mass radius of the galaxy, a proxy of its stellar size. Therefore, unless otherwise stated, $\MSTAR$ refers to this definition. As for gas mass, we include all gas gravitationally bound to the galaxy, without imposing any radial restriction, since a noticeable fraction of the gas may extend beyond $2\times \rhStar$.

We note that this is not directly comparable to any observational measurement \citep[see e.g.][for a discussion on this]{Pillepich2017}. However, firstly, this operational choice is well defined for both centrals and satellite galaxies, i.e. even in the case of galaxies that may undergo stripping. And secondly, for the purposes of this paper, the exact definition of galaxy stellar mass is not important, so long as we compare galaxy samples from the simulation with consistent definitions.

In certain instances, we also characterize central and satellite galaxies via their total mass or dynamical mass i.e. the summed mass of all their gravitational-bound stars, gas, dark matter and SMBHs within a distance of $<2\times \rhStar$, which we denote by $\MDYN$.

For some measurements, however, we use the summed mass of all gravitational bound stars or gas, which we denote by $M^{\mathrm{allgrav}}_{*/\mathrm{gas}}$.

\subsection{Measurements of star-formation activity}
\label{sec:sfr}

In the following Sections, we characterize the star formation activity of simulated galaxies via their star formation rates (SFRs), distances from the star forming main sequence and quenched fractions. Additionally, we distinguish between global i.e. galaxy-wide SFRs and SFRs measured within sub-components of a galaxy's body, as described below.

\subsubsection{Global SFRs}
\label{sec:globalsfr}
As a fiducial measure of the SFR of a galaxy, we choose to use the galaxy-wide `instantaneous' SFR directly available from the simulation catalogs: this is the sum of the SFRs of all gas cells that are gravitationally bound to a galaxy, irrespective of distance. For the case of jellyfish galaxies, in this way we cover both the SFR in the disk, or more generally main body, as well as in the tail.

Again, such a SFR metric is not a priori comparable to any observational measurement. However, \citealt{Donnari2019, Donnari2020a} have shown with TNG galaxies that the differences between this instantaneous measure and those based e.g. on averaging the SF over the last $10-1000$ Myr are completely negligible at all redshifts studied here, both in terms of the locus of the main sequence and in terms of quenched fractions.

In Appendix~\ref{appendix}, we show that the results of this paper are unchanged if we use two alternative estimates of the global SFR of a galaxy: 1) by restricting the SFR to its inner regions, i.e. by including only the contribution of gas cells within $<2\times \rhStar$, and 2) by also including those gas cells that may not be gravitationally bound to a satellite at the time of inspection but were bound at infall (procedure explained in Section \ref{sec:InfallCells}).

It should be noted that, given the finite mass resolution of the simulation, there is a minimum resolvable SFR for any galaxy. For TNG50 this is about $\sim 10^{-5}\,\MSUN$ yr$^{-1}$ at $z=0$. To handle this issue, we randomly assign an SFR value in the range of $10^{-6} - 10^{-5}\,\MSUN$ yr$^{-1}$ to those galaxies (or portions thereof) with SFR below the resolution limit of the simulation and that would otherwise have SFR = 0.

\subsubsection{Star forming main sequence and definition of quenched}
\label{sec:sfms}
To construct the star forming main sequence (SFMS) of TNG50 galaxies at any given time, we measure the mean SFRs of {\it star-forming} galaxies at the corresponding redshift in bins of galaxy stellar mass.
In turn, to distinguish between star-forming and quenched or green valley galaxies, we use the `Star Formation Activity Flags' and definitions of \citealt{Pillepich2019}, who identify the SFMS with a recursive method (whereby SFR is within $<2\times \rhStar$). 
In practice, star-forming galaxies have a logarithmic distance from the SFMS of $\Delta$log(sSFR)$>-0.5$ whereas all other galaxies could be called either green valley or quenched.

In the following, we also quantify the fraction of quenched galaxies to the total number of galaxies in bins of galaxy stellar mass. In that case, we define as quenched all galaxies with sSFR $\le 10^{-11}$ yr$^{-1}$, as is typically done in the observational community and suitable at low redshift \citep[see e.g.][for a discussion]{Donnari2020a}.

\subsubsection{SF in the main body vs. tails}\label{sec:distbodytail}
As we are interested not only in the global SF activity in jellyfish, but also where SF may occur and especially whether it occurs in the stripped gas tails, we divide all jellyfish galaxies into a body part and a tail component and measure their SFRs individually. To do so, we first define a distance $R_{\mathrm{dist}}$ to distinguish between disk and tail. Every gas cell closer to the position of the galactic center than $R_{\mathrm{dist}}$ is then counted as being part of the galaxy's body, while every gas cell farther away is counted in the tail. The simplest choice for $R_{\mathrm{dist}}$ is to use $2\times \rhStar$; however, for a small number of galaxies, this choice can lead to an underestimation of the extent of their body component and thus contaminate the tail component. To mitigate this issue, we instead choose to define $R_{\mathrm{dist}}=\mathrm{max}[2\rhStar,R_\mathrm{body}]$, where $R_\mathrm{body}$ is defined as follows.

We first define a vector $\vec{x}_\mathrm{long}$, which points from the galaxy's center to the gas cell (which is gravitationally bound to the galaxy) with the largest distance from the center. We construct a Cartesian coordinate system with its origin at the galactic center and its x-axis defined by $\vec{x}_\mathrm{long}$. For gas cells with an x-coordinate within $\pm\mathrm{r}_{1/2, *}$, we then calculate their distance $\mathrm{d}_\mathrm{perp}$ perpendicular to $\vec{x}_\mathrm{long}$ and use the largest such distance as $R_\mathrm{body}$.

\begin{figure}
    \includegraphics[width=\columnwidth]{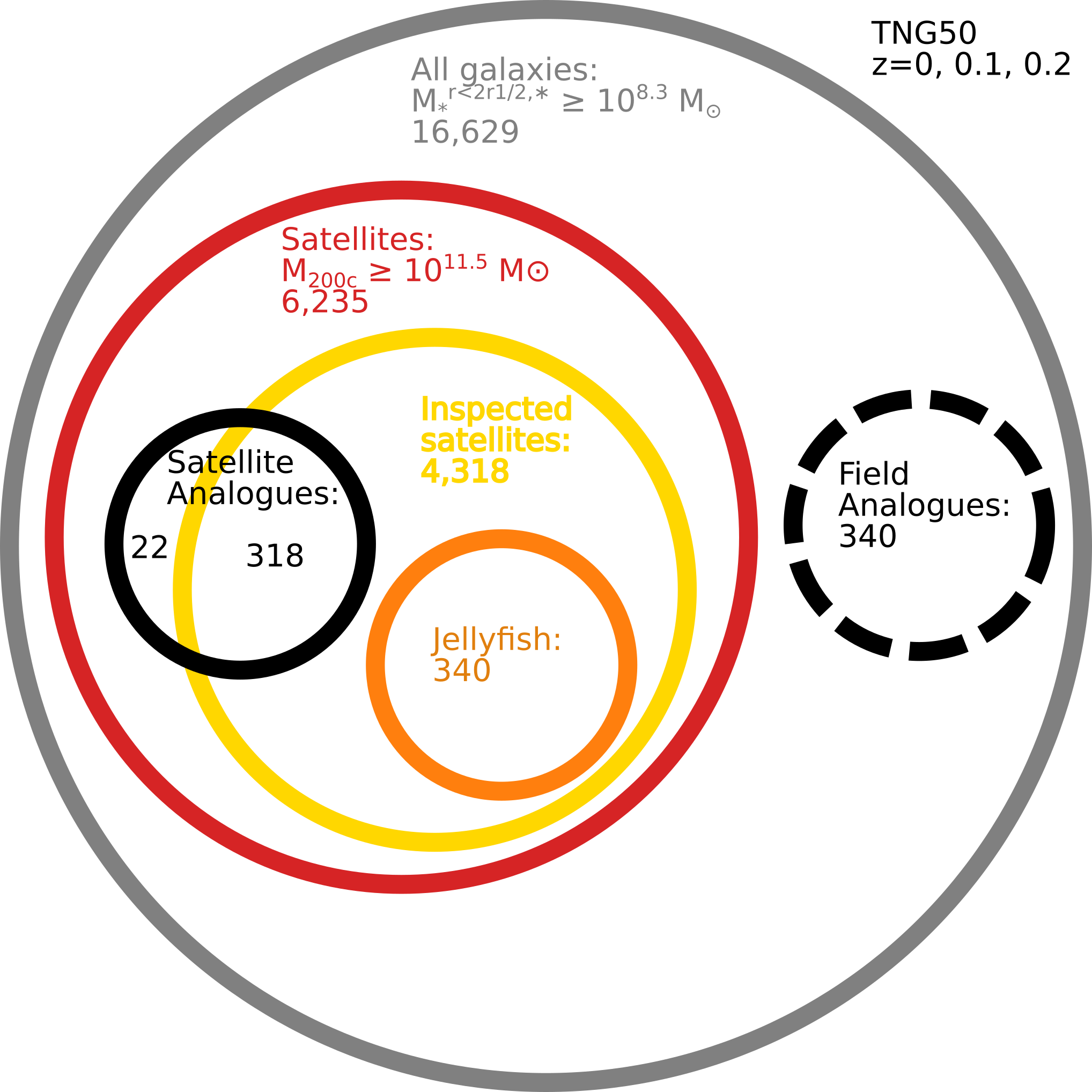}
    \caption{{\bf Schematic overview of the samples of TNG50 simulated galaxies studied in this paper, and their interconnection}. The numbers of objects refer to the case of the combined redshifts of $z=0$, $0.1$ and $0.2$. Throughout, we focus on galaxies with stellar mass $\gtrsim 2\times10^8\,\MSUN$ at $z<1$.}
    \label{fig:ovsample}
\end{figure}

\begin{table*}
    \centering
    \begin{tabular}{lcccccccc}
    \hline
    Snapshot \# & 99 & 91 & 84 & 78 & 72 & 67 & 59 & 50 \\
    z & 0.0 & 0.1 & 0.2 & 0.3 & 0.4 & 0.5 & 0.7 & 1.0 \\
    \hline
    All galaxies & 5635 & 5529 & 5465 & 5368 & 5274 & 5187 & 4971 & 4755\\
    All satellites & 2157 & 2064 & 2014 & 1952 & 1924 & 1849 & 1708 & 1472\\
    Inspected satellites & 1417 & 1434 & 1467 & 1488 & 1533 & 1501 & 1495 & 1423\\
    Jellyfish & 118 & 107 & 115 & 103 & 105 & 94 & 67 & 71\\
    \hline
    \end{tabular}
    \caption{{\bf Number of TNG50 galaxies considered in this work}, split in progressively smaller and more specific sub samples, as described in Section~\ref{sec:samples} and at all studied redshifts. The bottom rows give the number of jellyfish galaxies simulated within TNG50 and visually inspected in the Cosmological Jellyfish Zooniverse project.}
    \label{tab:Jelnumbers}
\end{table*}

\subsection{Galaxy selection, TNG50 jellyfish, and control samples}
\label{sec:samples}

In this paper, we focus on galaxies selected from the TNG50 volume at various cosmic epochs and that satisfy certain stellar mass, gas fraction, and satellite-vs-central criteria. In practice, the latter are dictated by choices that were in turn adopted for the visual classification of simulated satellites within the Cosmological Jellyfish Zooniverse project (see Section~\ref{sec:zooniverse} and below). 

In the following we describe how we select and how we divide TNG50 galaxies into subsamples that we contrast throughout. These are visualized in \cref{fig:ovsample} for a selection of redshifts combined together. Overall, we focus on TNG50 galaxies in simulation snapshots at the following redshifts: $z=0$, $0.1$, $0.2$, $0.3$, $0.4$, $0.5$, $0.7$ and $1$.\\

\textit{\textbf{All galaxies in this study:}} Throughout this paper, we only consider TNG50 galaxies above a certain stellar mass, to ensure that each be composed of at least a few thousand stellar particles: namely, $\MSTAR\geq10^{8.3}\,\MSUN$. Furthermore, we only consider \textsc{Subfind} objects of cosmological origin by using the `SubhaloFlags' \citep{Nelson2019b}. These criteria return the largest TNG50 galaxy sample of reference in this study, which is referred to as `All galaxies' and which is represented in grey in \cref{fig:ovsample}. \\

\textit{\textbf{Satellites:}} Throughout this work, we call `Satellites' all galaxies that are not the central of their FoF halo and whose host halo mass exceeds $\MHOST \geq 10^{11.5}\,\MSUN$. $\MHOST$ is the total, i.e. summed over all associated particles and cells, mass of the halo enclosed in a sphere with average density 200 times the critical density of the universe at the time of consideration. They form a subset of the `All galaxies' sample and are depicted in red in \cref{fig:ovsample}.\\

\textit{\textbf{Inspected satellites:}} We restrict the search of jellyfish galaxies in the context of the Cosmological Jellyfish Zooniverse project (Section~\ref{sec:zooniverse}) to TNG50 satellites only, and to those that still contain some gas at the time of inspection. This makes sense, in that there is no point in looking for gaseous tails if satellites have vanishing gas mass. In particular, following \citealt{JelZinger}, the sample of `Inspected satellites' (yellow in \cref{fig:ovsample}) is a sub-sample of `All satellites' and fulfils the following selection criteria:
\begin{itemize}
    \item they are satellites according to \textsc{Subfind}, i.e. are not the most massive galaxy of a FoF group;
    \item they have a stellar mass of $\MSTAR>10^{8.3}\,\MSUN$;
    \item they have a gas fraction of $\Mallgravgas/\MSTAR >0.01$.
\end{itemize}

\textit{\textbf{Jellyfish:}} In this paper, we call `Jellyfish' all galaxies among the `Inspected satellites' that received a weighted score of $\geq0.8$ according to the visual classifications of the Cosmological Jellyfish Zooniverse project. We thus use a definition identical to the one used in \citealt{JelZinger}. Please refer to this paper for a detailed discussion on different scoring and weighting schemes. These galaxies are depicted in orange in \cref{fig:ovsample}.\\

In order to assess whether jellyfish galaxies exhibit on average elevated or not SF activities (the goal of this paper), we compare their properties to two control samples constructed as follows.\\

\textit{\textbf{Satellite analogues:}} We first define a control sample of satellites that are analogous to the jellyfish but are not classified as such. These are chosen among `All satellites' to be the euclidean nearest neighbours to a jellyfish galaxy in a phase space of total galaxy stellar mass, gas-to-stellar mass fraction $\Mallgravgas/\Mallgravstar$ (which is limited to be strictly $>0$), and host mass $\MHOST$, with all dimensions normalized such that the values lie between 0 and 1 to ensure an equal weighting of all dimensions. Here the galaxy stellar and gas masses include all the gravitationally-bound stellar particles and gas cells, respectively. If they were inspected, which is the case for most of them, as shown by Fig.~\ref{fig:ovsample}, these satellites have a score smaller than $0.8$. Repetition is permitted for this comparison sample, i.e. one galaxy may be the analog to more than one jellyfish, however, about 79\% of the analogues are unique galaxies. The `Satellite analogues' sample (black in \cref{fig:ovsample}) contains exactly the same number of objects as the jellyfish sample, at each redshift. Importantly, jellyfish and their satellite analogues have the same exact distributions of galaxy stellar mass, gas content and host mass, which are all known to affect the SFR of satellite galaxies \citep[see e.g.][for TNG galaxies]{Donnari2020, Joshi2021}. Hence, if jellyfish and their satellite analogues have systematically different SFRs, we can more easily ascribe such difference to the jellyfish nature of the former rather than to other known internal or environmental properties.\\

\textit{\textbf{Field analogues:}} Similarly to the Satellite analogues, we compare Jellyfish with a population of TNG50 field galaxies. These include galaxies that are centrals of their FoF halo, are hosted by halos with $\MHOST<10^{11.5}\,\MSUN$, and are the nearest neighbours of jellyfish galaxies in a 2D phase space of total stellar mass and gas fraction (again limited to be $>0$ and normalized as above), at any given time. Here again we allow for repetition, and about 47\% of the galaxies are unique. This subsample is disjunct from the `All satellites' sample (see black dashed in \cref{fig:ovsample}). Again, we construct this control sample to ensure that any effect that we may be seeing for jellyfish galaxies is driven by the latter being jellyfish instead of being due to differences in other galaxy properties. \\

The number of TNG50 galaxies that we consider in this paper and that enter in each of the samples defined above at each considered redshift are given in \cref{tab:Jelnumbers}.

\subsection{Tracking unique jellyfish across cosmic time} \label{sec:tracking}

While a majority of this paper's analysis focuses on galaxy populations inspected at fixed redshift, in Section~\ref{sec:SFRbursts} we also follow jellyfish galaxies along their evolutionary history. In fact, based on the selection criteria adopted for the Cosmological Jellyfish Zooniverse project, frequently an individual galaxy was inspected multiple times at different epochs in cosmic history along its evolutionary track. Here, we follow the methodology developed and described by \citealt{JelRohr} (Section 2.3), whereby we connect the galaxies that were inspected at multiple times using the {\sc sublink\_gal} merger trees \citep{Rodriguez-Gomez2015}: from the $53,610$ satellites inspected at tens of snapshots in TNG50, there are 5,023 unique galaxies (or branches), and we follow their evolution across cosmic time.

\begin{figure*}
\centering
\includegraphics[width=1\textwidth]
{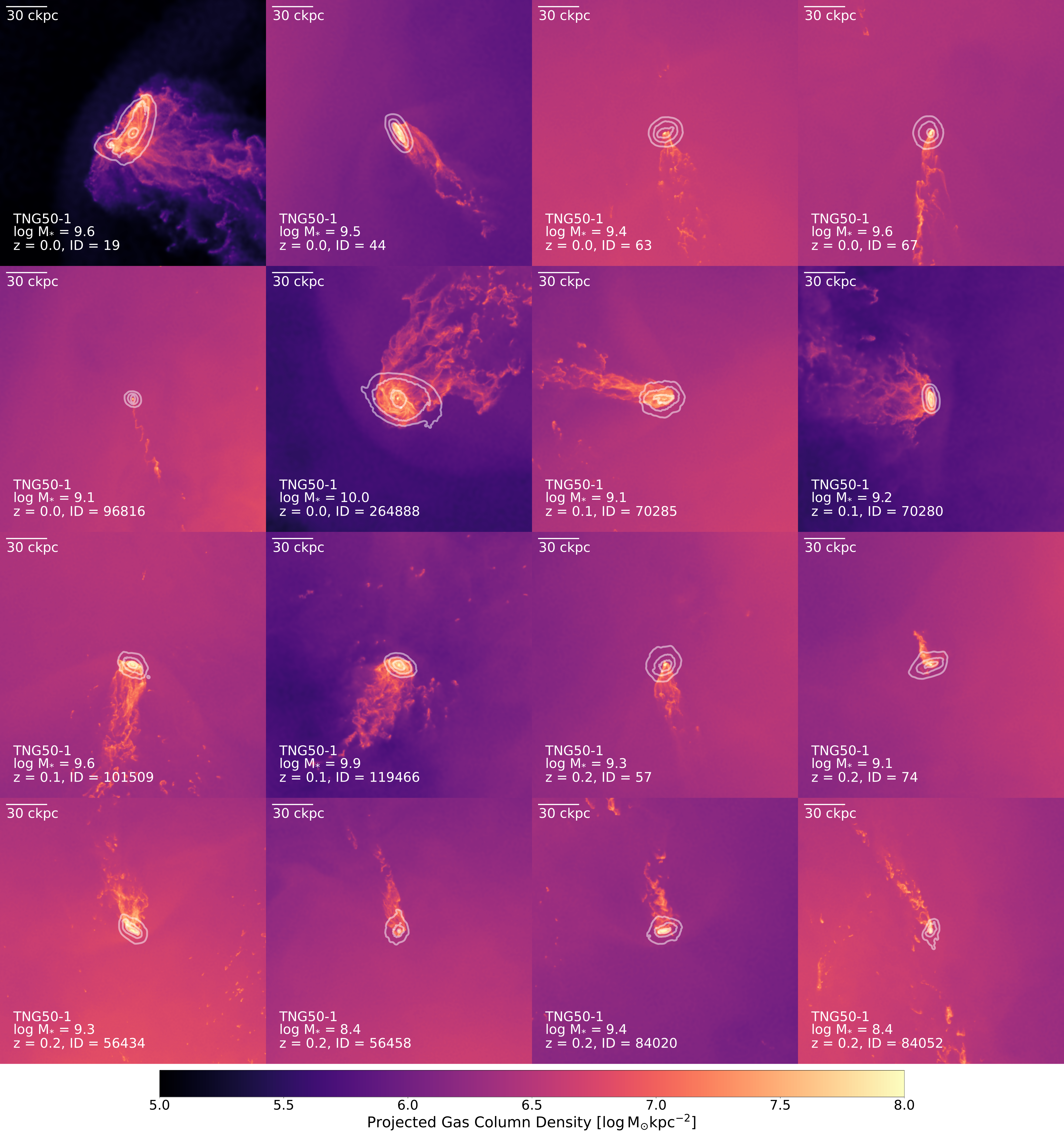}
\caption{{\bf Projected gas column density in a few example jellyfish galaxies from the TNG50 simulation.} These galaxies are selected among those with non-vanishing star formation in the ram pressure-stripped tails. Overlaid are stellar mass surface density contours; the contours indicate 60, 70 and 80 per cent of the peak stellar mass surface density. The extent of star-formation in these galaxies is shown in Fig.~\ref{fig:maps_sfr}.}
\label{fig:maps_column}
\end{figure*}

\begin{figure*}
\centering
\includegraphics[width=1\textwidth]
{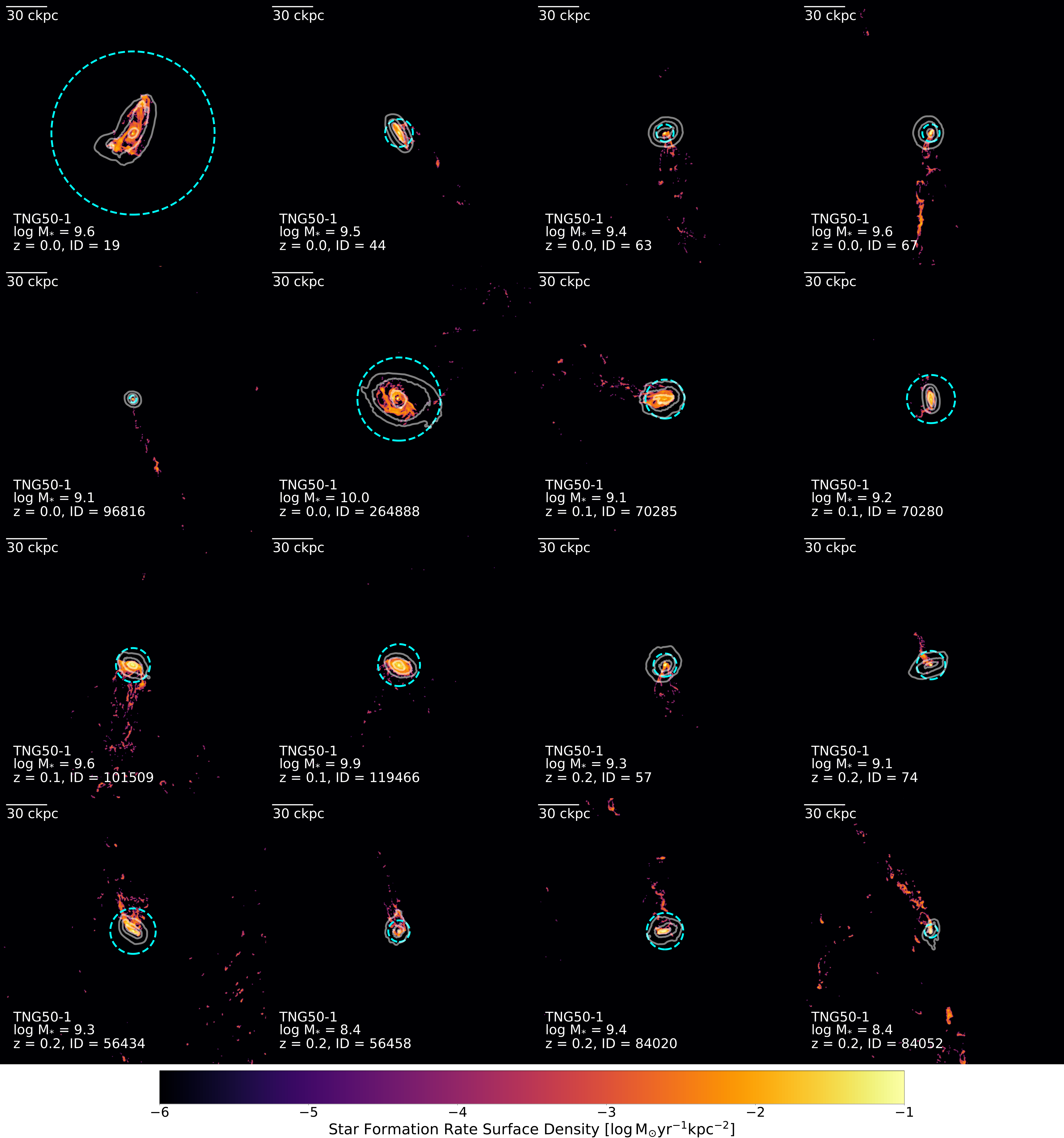}
\caption{{\bf SFR surface density in a few example jellyfish galaxies from the TNG50 simulation.} These are the same galaxies as shown in Fig.~\ref{fig:maps_column}, here showing the SFR within all gas around the galaxy. In turquoise we indicate the sphere (radius $R_{\rm dist}$) within which we count the galaxy's gas as being part of the galaxy body; see Section \ref{sec:distbodytail} for details. Gas (and therefore SF) outside the turquoise circle is part of the galactic tail. At least within the TNG model, the bulk of the gas in the tails is not star forming and the tails should appear much less extended and massive in any proxy of SFR (e.g. H$\alpha$).}
\label{fig:maps_sfr}
\end{figure*}

\section{Results with TNG50 jellyfish}

Based on the sample selections and the visual inspection procedures described above, the TNG50 simulation returns a total of 4144 jellyfish galaxies at $z\le2$ across all the 37 output snapshots. This sample size is more than an order of magnitude larger than that of any observational survey targeted at jellyfish galaxies, as well as larger than the sample size in any simulation-based study we are aware of. In this paper, we focus on $z\lesssim1$ and on a smaller set of available snapshots for a jellyfish population of 780 objects at various cosmic epochs (Table~\ref{tab:Jelnumbers}).

A selection of the TNG50 simulated jellyfish galaxies is shown in Figs.~\ref{fig:maps_column} and \ref{fig:maps_sfr}, in projected gas mass column density and SFR surface density, respectively. In the former, all gas associated with the satellite and its host is depicted, irrespective of temperature or phase. It is apparent that jellyfish galaxies, despite undergoing RPS, can be gas rich even in the main galaxy body. A few of them move supersonically through the ambient medium, producing majestic bow shocks (see Fig. \ref{fig:maps_column}, ID 19, 44, 264888 and 101509 or \citealt{Yun2019a}). On the other hand, most of their gas, in the bodies or tails, is not necessarily star forming, as the corresponding maps of Fig.~\ref{fig:maps_sfr} show, at least according to TNG50. In the following sections, we quantify the SF activity of TNG50 jellyfish galaxies. However, we point out that for the jellyfish galaxies identified in TNG50, RPS acts directly on their cold gas, with the long-lived jellyfish tails originating mostly from the cold ISM of satellite galaxies \citep{JelRohr}.

\begin{figure*}
\centering
\includegraphics[width=0.3\textwidth]{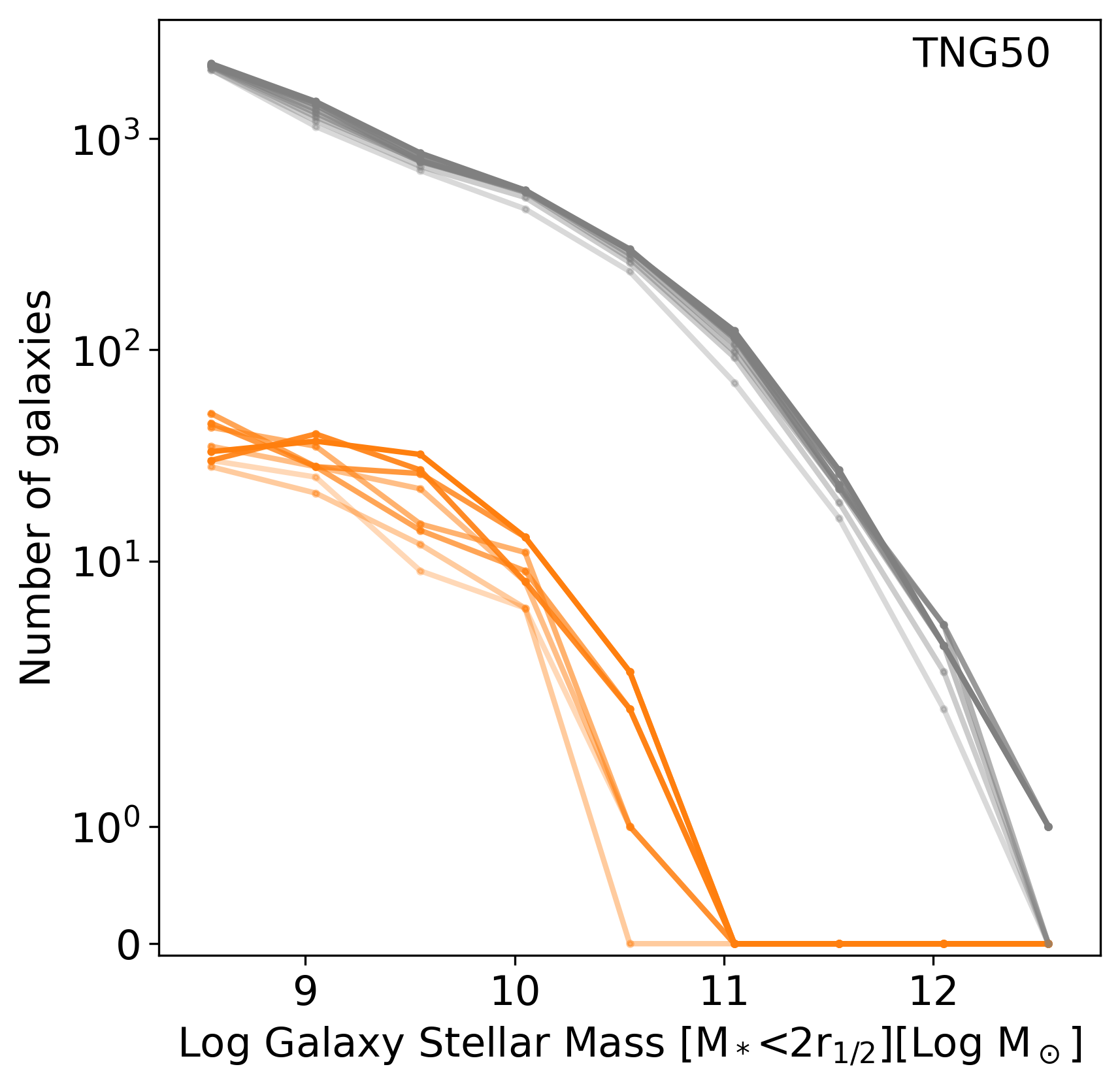}
\includegraphics[width=0.3\textwidth]{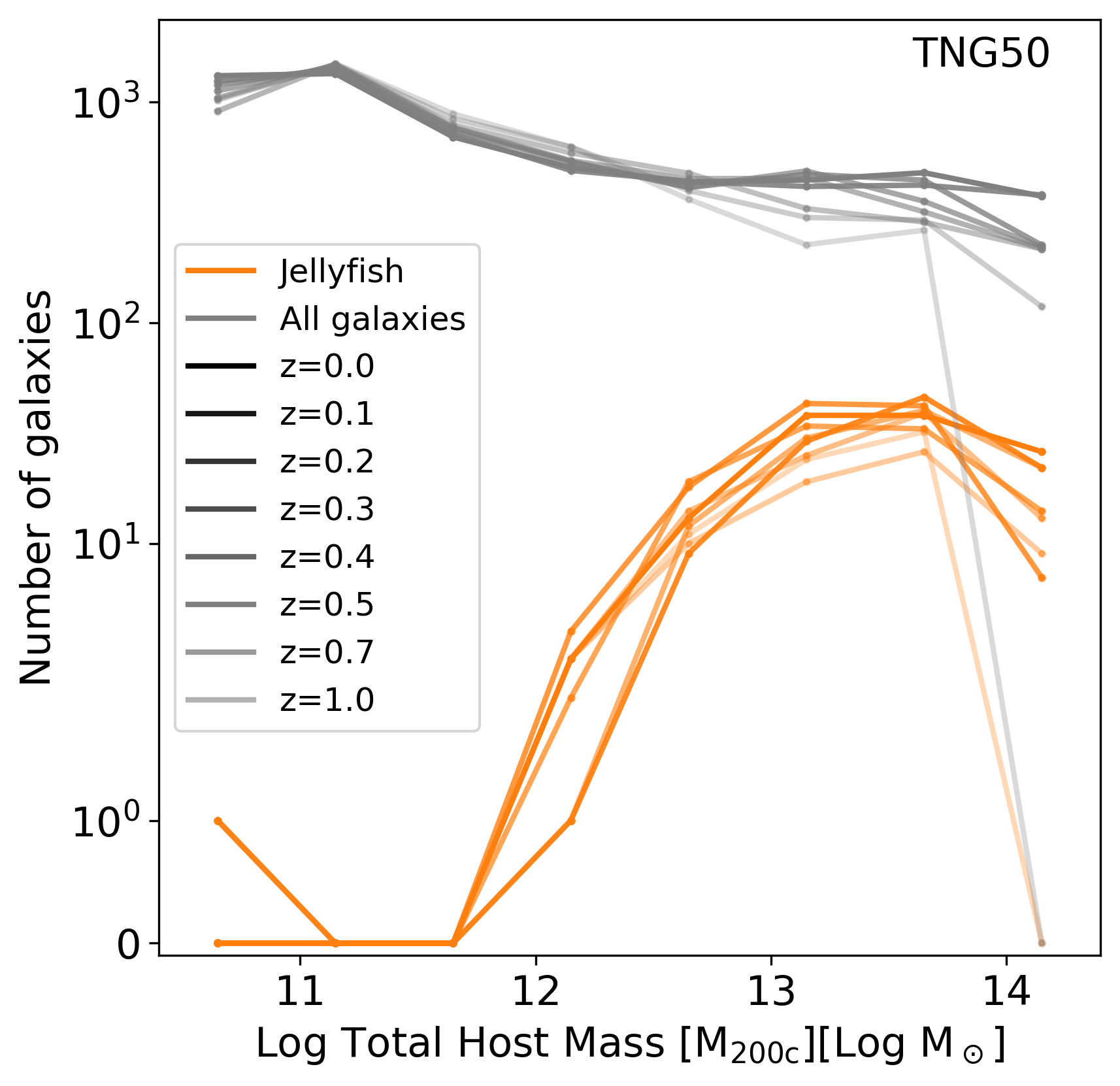}
\includegraphics[width=0.318\textwidth, trim={0cm, 0cm, 0.63cm, 0cm}, clip]{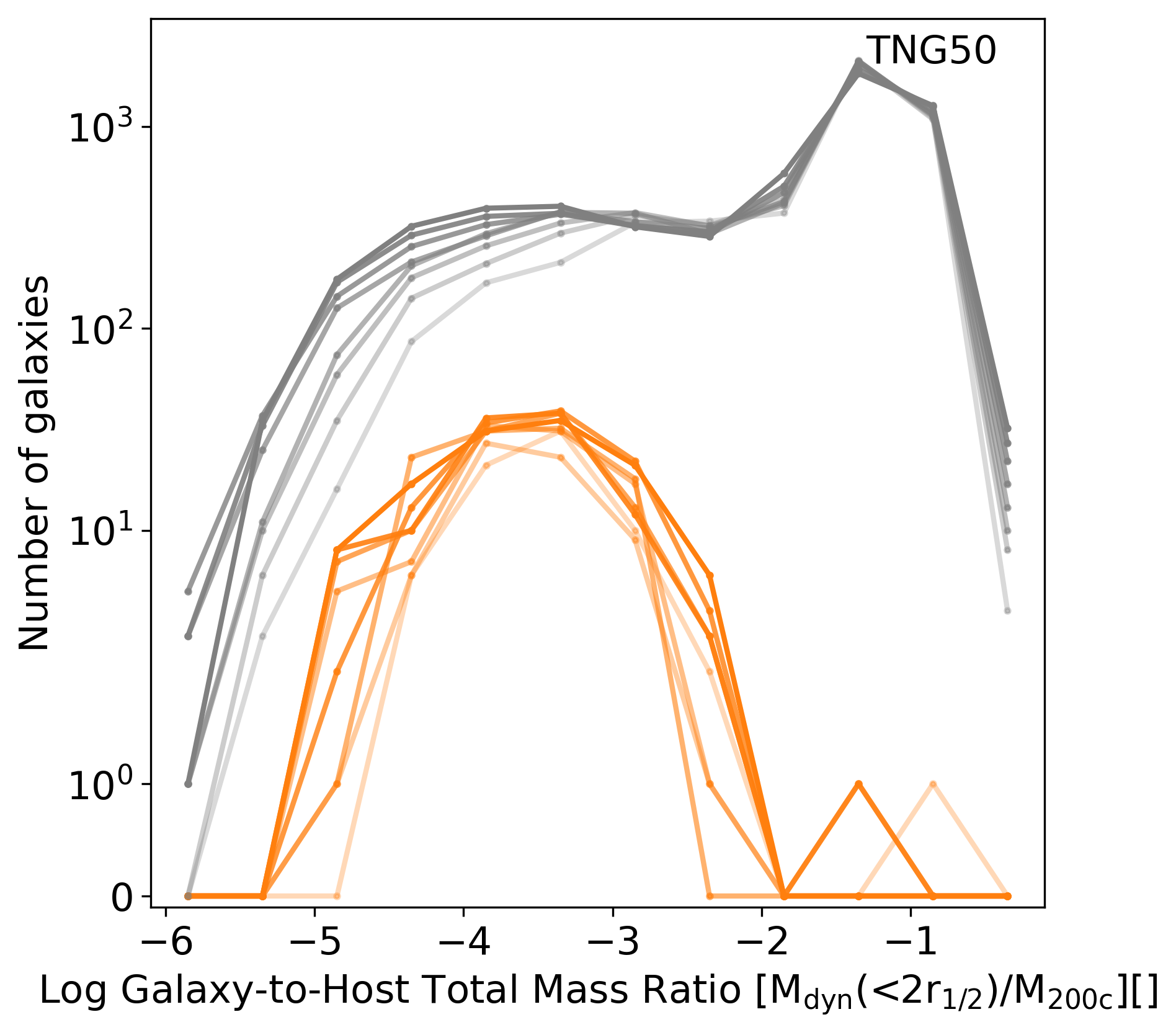}
\caption[Frequency of jellyfish galaxies in the galaxy parent sample.]{{\bf Number of all galaxies and jellyfish galaxies from the TNG50 simulation studied in this paper, as a function of stellar mass, host halo mass and satellite-to-host mass ratio, at each redshift in our sample}. The frequency of jellyfish galaxies decreases with stellar mass and increases with host mass, indicating that jellyfish galaxies are more abundant in more massive host halos and at lower stellar masses. However, most jellyfish galaxies are found at a satellite-to-host mass ratio of $10^{-4}$ to $10^{-3}$ and not at the lowest ratios.} 
\label{fig:Demograph}
\end{figure*}

\subsection{Demographics and properties of TNG50 jellyfish galaxies} \label{sec:demographics}

In order to understand the star-formation activity of the jellyfish galaxies in TNG50, it is first important to describe their demographics and the environments they are found in.

In fact, as shown for TNG100 satellites by \citealt{Yun2019a} and for TNG50 and TNG100 by \citealt{JelZinger}, jellyfish are more frequent, in comparison to the inspected sample, at cluster-centric distances between $0.5$ and $1\,\mathrm{R}_\mathrm{vir}$, in more massive hosts and at smaller satellite masses, and they typically orbit supersonically. All this holds in comparison to satellites selected for inspection, and hence also still containing some gas.

It is in fact generally known that jellyfish (and more generally, satellite) stellar mass plays an important role in determining any environmental effects, as does the host halo mass. Therefore, to set the stage and properly characterize our galaxy samples, we first examine the distributions of the galaxy stellar mass, host halo mass, and the ratio of the \emph{total} galaxy mass to host mass for jellyfish galaxies, and compare them to those of the full sample of `All galaxies' in Fig.~\ref{fig:Demograph}.

The left panel of Fig.~\ref{fig:Demograph} shows the stellar mass function of the `Jellyfish' and `All galaxies' samples,
separately at all redshifts considered in this analysis. Firstly, this panel shows that TNG50 returns a wide range of galaxy stellar masses, from our mass limit of a few $10^8\,\MSUN$ up to massive galaxies of $\sim10^{12}\,\MSUN$ in stars (grey curves). The lack of jellyfish galaxies with $\MSTAR>10^{11}\,\MSUN$ (orange curves) is largely due to the fact that we required the inspected galaxies to be satellites, which in turn severely restricts the number of inspected massive galaxies and therefore jellyfish galaxies. Moreover, at stellar masses $>10^{10.5}\MSUN$, the feedback from SMBHs can remove large fractions of the galactic gas \citep{Zinger2020}, therefore lowering the possibility of the galaxy being a jellyfish galaxy -- see a preliminary discussion on high-mass satellites and their chances of appearing as jellyfish in \citep{JelZinger}.

Secondly, there is little redshift evolution in the mass functions for `All galaxies', either in terms of shape or normalization, except at the highest masses (i.e. $\MSTAR>10^{11}\MSUN$), where there is a mild increase in the number of galaxies with decreasing redshift, as would be expected since these are dominated by central galaxies that continually grow in stellar mass. In the case of the jellyfish galaxies, the stellar mass functions have shapes similar to the `All galaxies' mass functions at all redshifts. Furthermore, the normalization of the mass functions shows mild evolution at best, with more jellyfish galaxies being found at lower redshifts, although this trend is not clear at all masses. 

Thirdly, when compared to the overall galaxy population at similar stellar masses, jellyfish galaxies are clearly not frequent, representing less than a few percent of the whole galaxy population \citep[see also][and \citealt{JelZinger}]{Yun2019a}. Note that from Table \ref{tab:Jelnumbers}, the fraction of the overall `Inspected satellites' sample that are jellyfish galaxies notably increases from 5 per cent at $z=1.0$ to 8 per cent at $z=0$. This implies that the jellyfish galaxies represent an approximately constant (w.r.t. stellar mass, mildly increasing with redshift) fraction of all galaxies, except for the most massive galaxies i.e. $\MSTAR \gtrsim 10^{10.3}\,\MSUN$. The decreased fraction of jellyfish at the highest stellar masses is again largely due to the fact that jellyfish are satellite galaxies and very massive satellites ($\MSTAR\gtrsim10^{12}\,\MSUN$) can generally be hosted only by very massive halos ($\MHOST\sim10^{15}\,\MSUN$), which are absent in the TNG50 volume (see also in the next paragraphs). Furthermore, the increased gravitational strength in more massive galaxies binds gas to the galaxy more efficiently and therefore hinders stripping. Finally, massive galaxies are affected by SMBH feedback, at least in IllustrisTNG, irrespective of whether they are centrals or satellites \citep{Donnari2020}, which in turn expels large amounts of gas from their inner regions, possibly affecting the chances of them being jellyfish \citep{Terrazas2020, Zinger2020} (see also next Sections).

In the middle panel of Fig.~\ref{fig:Demograph}, we show the distribution of the host masses $\MHOST$ for the two samples. As a reminder, the grey curves (`All galaxies') include both centrals and satellites. For these, the underlying host halos span the entire available mass range, with halos of $10^{11}$ to $10^{12}\,\MSUN$ being more frequent, as is to be expected for a volume limited sample in a $\Lambda$CDM scenario. On the other hand, jellyfish galaxies are more frequent in halos of mass $\MHOST = 10^{13-14}\,\MSUN$. We find little dependence on redshift for either sample, except perhaps at the highest host masses, which is likely due to the lower number of available massive hosts at higher redshifts. The stark difference in the shapes of the distributions of the host mass functions between the `All galaxies' and `Jellyfish' samples can be largely attributed to two factors, beyond the fact that Jellyfish are, by construction and nature, satellites: (i) more massive hosts cause higher infall velocities, which in turn cause more severe stripping and (ii) more massive hosts have a higher ICM density \citep{Domainko2006}, again increasing the ram-pressure exerted on infalling galaxies. We see a decline in the number of jellyfish galaxies at very high host masses, again most probably caused by the decreasing number of hosts of such a high mass.

These results reveal a preferred combination of properties for the occurrence of jellyfish galaxies: low mass galaxies in high mass hosts. In fact, not only is the stellar mass of the galaxies an important factor, so is their total mass within the central regions, as this determines their resilience to RPS. We check this explicitly in the right panel of Fig.~\ref{fig:Demograph}, where we show the distribution of the ratios of the galaxies' \emph{total or dynamical} mass to their host mass. The `All galaxies' sample displays a double-peaked distribution, with a broad peak at $\MDYN/\MHOST \sim 10^{-5}-10^{-3}$ and a narrower peak at $\MDYN/\MHOST \sim 10^{-2}-10^{-1}$. The former is likely representative of satellites, whereas the latter is likely due to the centrals in the full sample. This also explains the time evolution seen at the lower-ratio regime, which is where the lowest-mass satellites in the highest mass haloes would be found, reflecting the evolution of the number of galaxies found in the most massive hosts, as seen in the middle panel of Fig.~\ref{fig:Demograph}. The `jellyfish' galaxies however exhibit a markedly different distribution with typical dynamical mass-to-host mass ratios in the range of $\MDYN/\MHOST \sim 10^{-5}-10^{-2}$ and a single peak at $\MDYN/\MHOST \sim 10^{-3.5}$. The second peak is missing because jellyfish galaxies are exclusively satellites. The decrease in number towards the lowest ratios is explained by the fact that these low ratios represent a combination of the highest mass hosts and lowest mass jellyfish galaxies. These combinations are rare, even in the 'All galaxies' sample. As with the `All galaxies', we find mild evolution with redshift in the number of Jellyfish since $z=1$, such that more jellyfish galaxies are found at later redshifts for all values of $\MDYN/\MHOST$, although there is significant noise due to low-number statistics at the high and low values. These results are qualitatively consistent with those of \citealt{Yun2019a}, although note that that study only considers satellites in hosts of mass $\MHOST\geq10^{13}\MSUN$, and with the findings in the companion papers by \citealt{JelZinger} and \citealt{JelRohr}.

\subsection{Star formation in TNG50 jellyfish, even in the tails} \label{sec:disk_tail}

After getting familiar with the demographics and environments of jellyfish, we can now concentrate on their star formation activity. In the following sections we show that, according to TNG50, despite being severely affected by RPS, jellyfish galaxies are typically star forming.

\begin{figure*}
\centering
\includegraphics[width=0.8\textwidth]{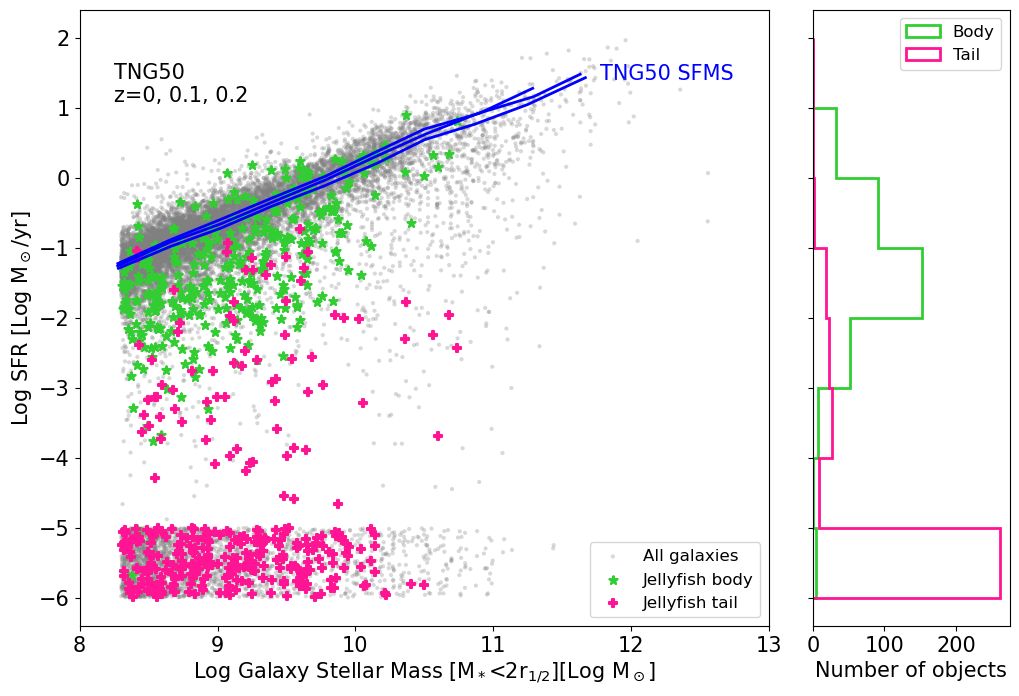}
\caption{{\bf Star formation in the bodies and tails of TNG50 jellyfish galaxies.} We show the SFR as a function of stellar mass for all TNG50 galaxies above a stellar mass of $10^{8.3}\,\MSUN$ (grey) and of jellyfish galaxies. The SFRs of jellyfish galaxies are shown separately for their body (green stars) and tail (pink pluses) components. The TNG50 SFMSs at the different considered redshifts are shown in blue. The SFMS is the locus of the mean SFR of the star-forming galaxies only, at the indicated redshifts: see text for detail. Galaxies with an SFR below $10^{-5}\,\MSUN/\mathrm{yr}$ are below the resolution limit of TNG50 and are assigned a random SFR value at the bottom of the figure. The majority, but not all, of the jellyfish tails have values of SFRs below the SFMS (see right panel), unlike the jellyfish bodies, which have significantly higher SFRs, but on average lower than the overall SFMS in TNG50. Only a few jellyfish galaxies have a SFR exceeding the SFMS; however, remarkably, one jellyfish galaxy exhibits SFR levels above the SFMS in the tail alone.}
\label{fig:sfr_bodytail}
\end{figure*}

This is quantified in Fig.~\ref{fig:sfr_bodytail}, where we show the SFRs of TNG50 jellyfish galaxies at $z=0, 0.1, 0.2$ (higher redshifts we examine in Section \ref{sec:SFRredshift} and \ref{sec:SFRbursts}) in the bodies and tails of the jellyfish galaxies, separately, as described in Section~\ref{sec:distbodytail}. In the main panel of Fig.~\ref{fig:sfr_bodytail} we present the global SFRs of `All galaxies' in TNG50 (grey dots) as a function of their stellar mass and the corresponding TNG50 SFMS (shown in blue, see Sections~\ref{sec:globalsfr} and \ref{sec:sfms}). Results are given for the combined galaxy populations at three different redshifts to gain a larger sample size; however the SFMS does not evolve appreciably during these epochs (blue curves). The SFR of jellyfish galaxies is shown divided into the contributions from their bodies (green stars) and RP-stripped tails (pink pluses), separately. Galaxies (or portions thereof) with SFRs below the resolution limit of TNG50 are assigned a random value between $10^{-5}$ and $10^{-6}\,\MSUN$ yr$^{-1}$. The right inset of Fig.~\ref{fig:sfr_bodytail} also shows the distribution of SFRs for the jellyfish bodies and tails.

It is apparent that TNG50 predicts star formation in the bodies as well as in the tails of jellyfish galaxies. Therefore, even though these galaxies undergo extreme RPS, they are not necessarily quenched and even have star forming regions in their tails. Such regions can be seen in Fig.~\ref{fig:maps_sfr}. However, the median SFR in the TNG50 jellyfish bodies is $3-4.8$~dex higher than in the tails: this is a significantly larger difference than the one of \citealt{Gullieuszik2020}, who find that the SFR in jellyfish tails compared to jellyfish bodies is reduced by a factor of $\sim5$, whereas it appears more in line with the numerical findings of \citealt{Kronberger2008}, whereby the SFR in jellyfish bodies is largely dominant over that in the tails. The body vs. tail difference in TNG50 increases with increasing stellar mass. More than three quarters of the jellyfish tails in our sample actually exhibit negligible, i.e. below the resolution limit, SFRs, while this is only true for about one per cent of the jellyfish bodies. 

Therefore, as can be appreciated by the comparison in the right panel of Fig.~\ref{fig:sfr_bodytail} and qualitatively by comparing the maps of Figs.~\ref{fig:maps_column} and \ref{fig:maps_sfr}, according to the TNG50 simulation, SF can occur in the RP-stripped tails of jellyfish galaxies, but typically at levels that are far lower than the SF in their main bodies. Furthermore, from the maps, it is clear that jellyfish tails detected in gas phases that trace SF are typically less pronounced, less extended or less broad than the tails across gaseous phases -- if they show tails at all. Still, it is interesting to note that a few systems at these low redshifts have SF in the tails at levels that are compatible with those of the SFMS, or even higher. How frequently this happens actually does depend on the specific method adopted to separate body and and disk vs. tails (Section~\ref{sec:distbodytail}). We have tried independent approaches and can confirm however that the overall picture remains qualitatively unchanged: TNG50 predicts jellyfish tails with ongoing star formation but typically this does not happen frequently and the SF in the main galaxy bodies always dominates.

\subsection{No enhanced population-wide star formation in jellyfish} \label{sec:global_sfr}

\begin{figure*}
\centering
\includegraphics[width=0.90\textwidth,trim={0cm, 0cm, 0cm, 0cm}]{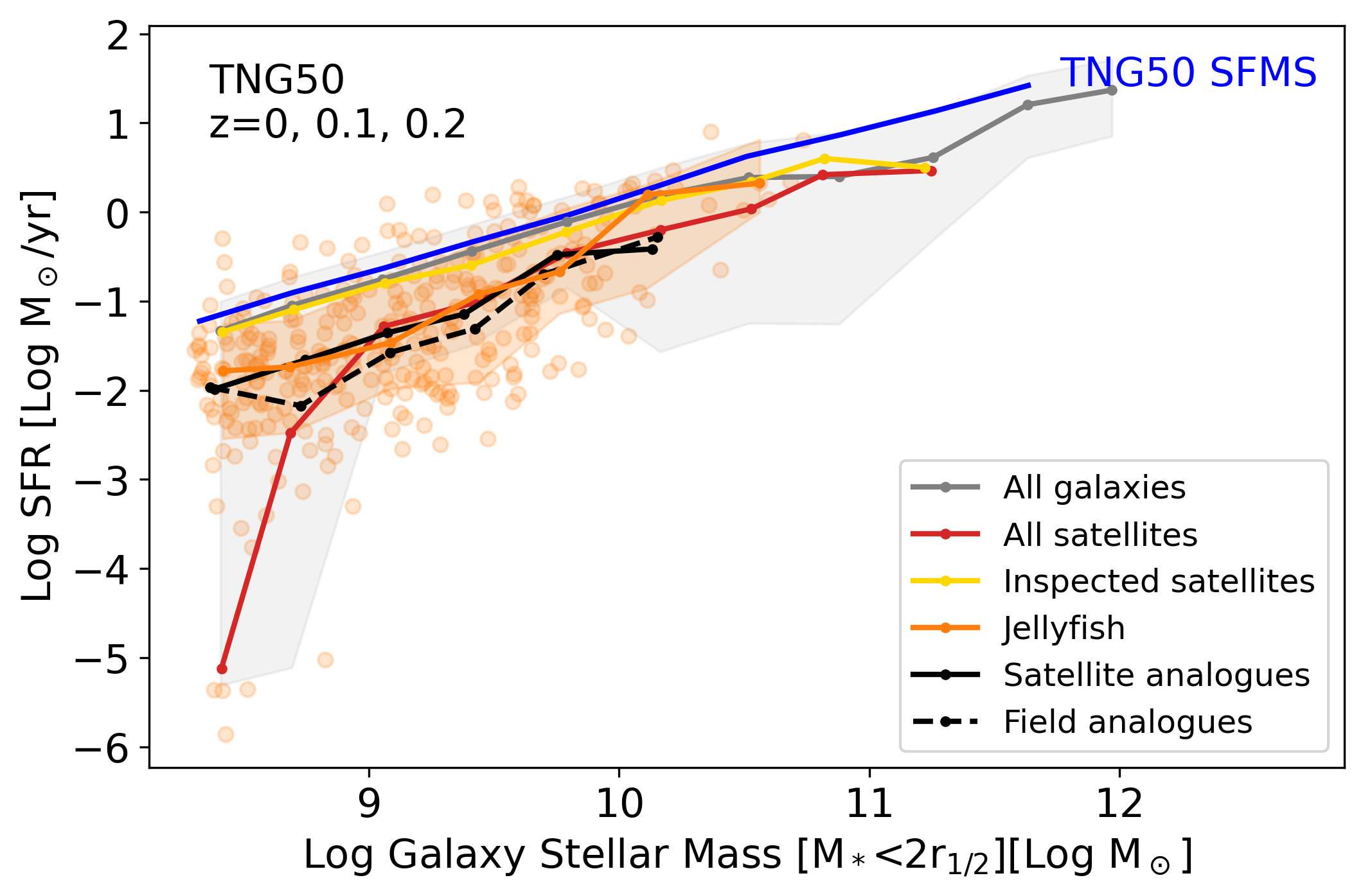}
\includegraphics[width=0.32\textwidth]{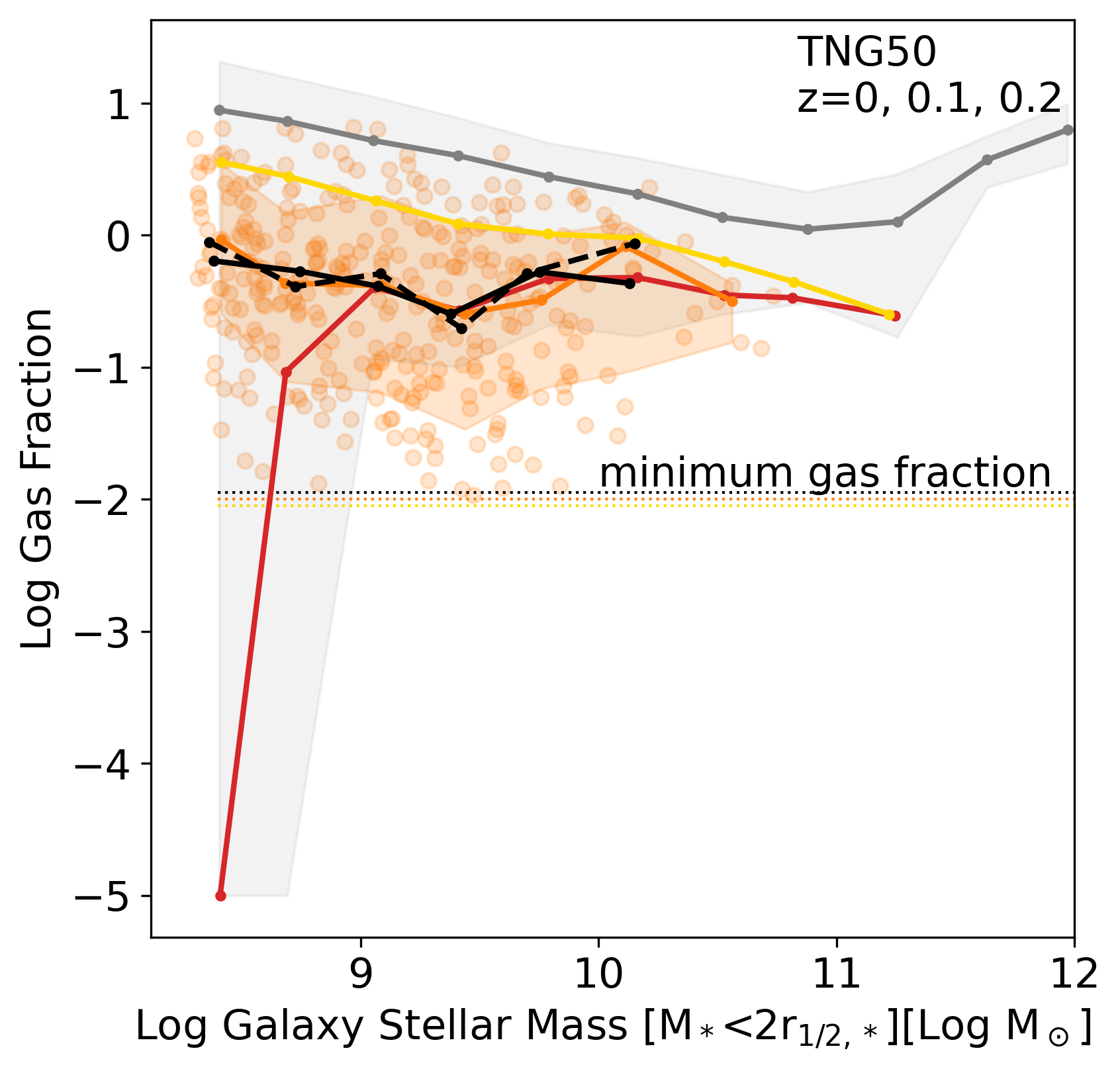}
\includegraphics[width=0.32\textwidth]{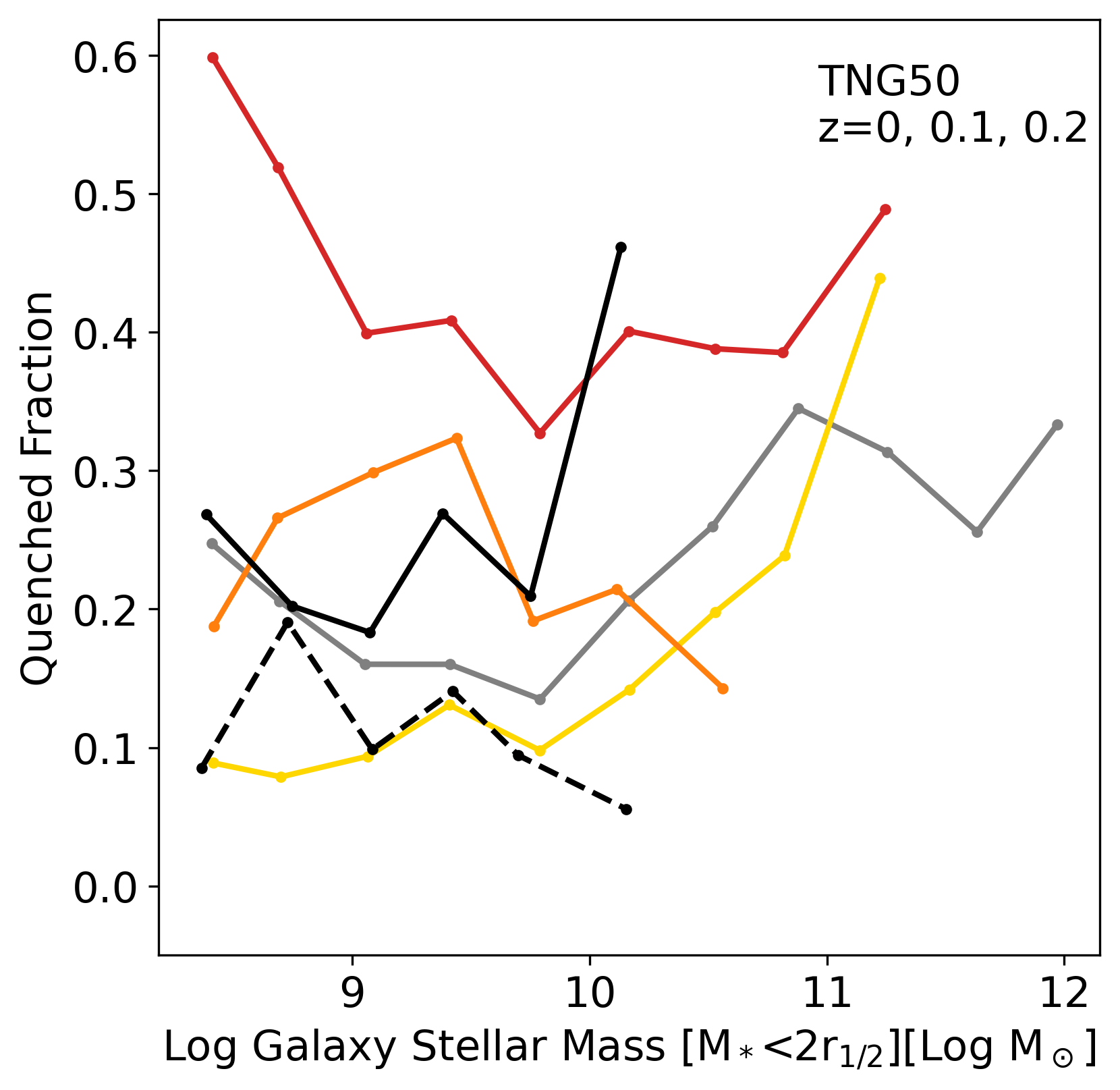}
\includegraphics[width=0.32\textwidth]{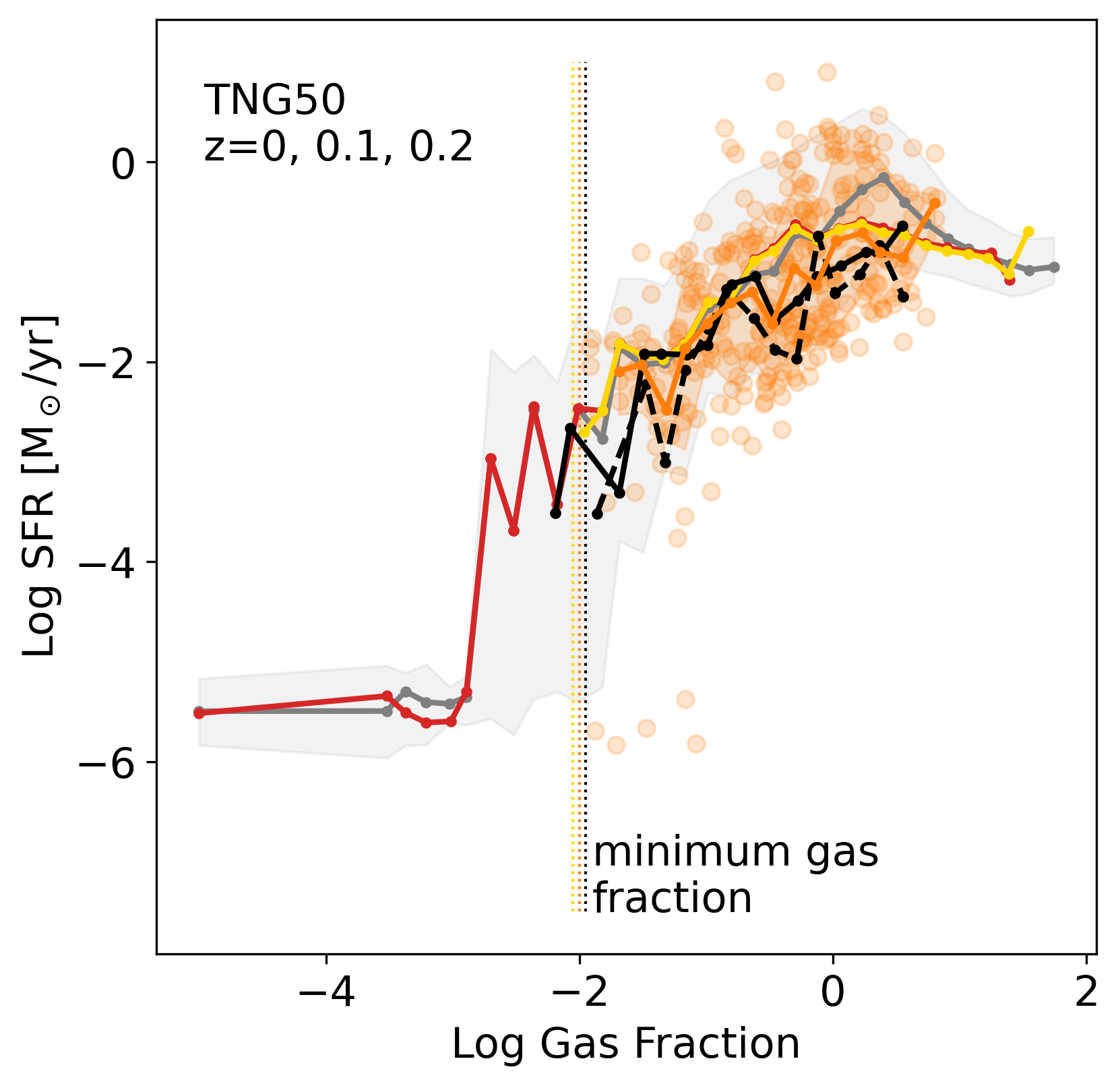}
\caption{{\bf Star formation activity of jellyfish galaxies in comparison to all TNG50 galaxies and control samples.} We show the median SFR as a function of stellar mass (top panel) for all galaxies (grey), all satellites (red), inspected satellites (yellow), jellyfish galaxies (orange) and their satellite (black) and field analogues (black dashed). The TNG50 SFMS is given in blue. We also show the median gas fractions (bottom left) and quenched fractions (bottom center) as a function of stellar mass and the median SFR as a function of gas fraction (lower right) for the same subsamples. Each bin contains at least 7 galaxies and bins with a lower number of galaxies are discarded. Bins of stellar mass have a width of $0.37$~dex; bins of gas fraction are $0.17$~dex wide. Shaded areas denote the 16th to 84th percentile for jellyfish galaxies and the `All galaxies' sample. The jellyfish galaxies show suppressed typical SFRs compared to our full TNG50 sample of galaxies and the `Inspected satellites' out of which they have been identified, but similar SFRs compared to  the `All satellites' sample and both control samples. Therefore, according to TNG50, there is no population-wide SFR enhancement in jellyfish. The suppressed typical SFRs are correlated with lower overall gas fractions and somewhat higher quenched fractions, but along with similar SFRs for a given gas fraction as any other galaxy.}
\label{fig:sfr}
\end{figure*}

From the above considerations, we can proceed by characterizing the SFR of jellyfish galaxies with a {\it global} measure (Sections~\ref{sec:globalsfr} vs. \ref{sec:distbodytail}), knowing that the latter is typically dominated by the SF that occurs in the main galaxy body rather than in the tails. Hence, we compare the global SFRs of the jellyfish galaxies with the various subsamples: all satellites, inspected satellites, satellite analogues, and field analogues, as described in Section~\ref{sec:samples} and Fig.~\ref{fig:ovsample}. We are thus able to distinguish subtle differences in both the selection effects (which also affect observational studies) and between satellites undergoing environmental effects but are not undergoing visually identifiable RPS and those that are, i.e. jellyfish. We aim to understand whether there is any overall population-wide enhancement or suppression of the SFRs of the TNG50 jellyfish galaxies compared to any of our samples.

In Fig.~\ref{fig:sfr} we therefore compare the median SFR as a function of stellar mass for the jellyfish galaxies to that of all other TNG50 samples described in Section~\ref{sec:samples} (main, top panel). Moreover and importantly, we interpret the top panel in view of the the gas fractions (bottom left) and quenched fractions (bottom middle) of galaxies as a function of stellar mass, and of the median SFR as a function of gas fraction (bottom right). In all cases, we include all TNG50 galaxies, i.e. even those with non-resolvable levels of SF (or gas mass), which are hence placed by hand at very small but non-vanishing SFR (gas mass) values and which also contribute to the SFR averages and medians.

\subsubsection{SFRs in jellyfish galaxies and other galaxy types} \label{sec:sfr_comparison}
In the main panel of Fig.~\ref{fig:sfr}, we quantify the median SFRs in bins of galaxy stellar mass of all TNG50 galaxy samples for redshift $z=0$, $0.1$ and $0.2$ combined. We use bins of $0.37\,$dex in stellar mass and discard those with fewer than 7 galaxies. Additionally, we plot the individual SFR values of the jellyfish as orange dots. We also indicate the 16th to 84th percentile in each bin as shaded areas for jellyfish and `All galaxies'.

The median SFR of `All galaxies' (grey, including SFing, green-valley and quenched galaxies) is very similar to the locus of the SFMS (blue, based exclusively on SFing galaxies), barring at the highest-mass end. This is expected and a good confirmation: we have checked that, at all considered masses, galaxies are more frequently centrals than satellites and centrals are known, in both observations and in the IllustrisTNG simulations, to have higher SFRs than satellites of the same mass (see Section~\ref{sec:intro} and references therein). At the high-mass end, the fraction of quenched galaxies increases both in the IllustrisTNG simulations and in reality and irrespective of central vs. satellite status \citep[e.g.][]{Donnari2019, Donnari2020}, lowering the average SFR. At the low-mass end ($\MSTAR\lesssim10^{10}\,\MSUN$), TNG50 and observed galaxies are typically star forming {\it unless} they are satellites, because they are affected by environmental processes \citep[e.g.][]{Donnari2020, Joshi2021}. However, in the grey curve, the latter are not the dominant population.

Again, as expected from both observations and simulations (see also Section~\ref{sec:intro}), satellite galaxies are affected by environmental processes and have on average lower SFRs than the whole galaxy population (red vs. grey curves), particularly below $10^{10}\,\MSUN$ (whereby these are more than $0.4-0.7$~dex below the SFMS). In fact, the median SFR of `satellite' galaxies experiences a sharp drop-off below a stellar mass of about $10^{9}\,\MSUN$ due to the large number of galaxies with a SFR below $10^{-5}\,\MSUN$ yr$^{-1}$. This is not the case for the `Inspected satellites' and their subsamples because, in order to identify jellyfish galaxies, we only considered galaxies with at least some gas (and hence more likely to have some star formation). 

It is also because of this selection effect due to the requirement of a minimum amount of gas that the `Inspected satellites' (yellow curve) exhibit similar SFR values as those of the general galaxy population (yellow vs. grey), i.e. {\it despite the fact} that they are all satellites. On the other hand, TNG50 jellyfish galaxies below $10^{10}\,\MSUN$ exhibit median SFRs that are $0.6-0.8$~dex lower than the SFMS (orange vs. blue) and $0.3-0.7$~dex lower than the satellites out of which they have been identified (orange vs. yellow). That is, although the jellyfish galaxies are all from the sample of inspected galaxies, their SFR is up to $0.7$~dex lower (see Section \ref{sec:gasfrac} for more explanation). 

These results show that, not only do the jellyfish galaxies have typically lower SFRs compared to the full sample of `All galaxies', which follows not only from the fact that the latter sample is dominated by central galaxies, but also that the jellyfish galaxies do not have a population-wide enhanced SF activity compared to other `Satellite' galaxies, except at stellar masses $\lesssim 10^{9}\, \MSUN$. In fact, we can go one step further and compare the median SFRs of the jellyfish galaxies to those of the two control samples of `Satellite analogues' and `Field analogues' to ensure than our results are not driven by differences in the stellar mass or gas fraction distributions, or host halo mass in the case of the former. The median SFR of `Satellite' (black) and `Field' (black dashed) analogues follow the median SFR of jellyfish galaxies closely; although `satellite analogues' actually show deviations up to $0.6$~dex, we think this is largely an effect of low number statistics. These results hence confirm that in TNG50, the SFRs of jellyfish galaxies do not differ significantly from satellite and field galaxies with similar masses and gas fractions. 

These TNG50 results are more in line with observational results that find no overall enhancement of SF activity in their identified jellyfish galaxies (see Sections~\ref{sec:intro} and \ref{sec:discussion}) and appear to be at odds with the observational findings of \citealt{Vulcani2018, Ramatsoku2020} and \citealt{Vulcani2020} based on the GASP survey. However, comparisons to such studies, especially with the latter two, are not trivial and require careful consideration of the adopted methods and quantities. We discuss these factors in more detail in Section \ref{sec:discussion_observations}. It should be noted however that, while the jellyfish galaxies have lower SFRs typically, it is apparent from both Figs.~\ref{fig:sfr_bodytail} and \ref{fig:sfr} that jellyfish galaxies with above-average SFRs do exist in TNG50. They are especially frequent for galaxies with $\MSTAR\sim10^{9-10}\,\MSUN$. Additionally, those jellyfish galaxies that are high in SFR are also high in gas fraction (see lower right panel of Fig.~\ref{fig:sfr}) and mostly have gas-to-stellar mass fractions $> 0.1$, as we show in the next section. 

\subsubsection{Gas fractions}\label{sec:gasfrac}
In the lower left panel of Fig.~\ref{fig:sfr} we quantify the typical gas fractions in the studied TNG50 galaxies as a function of stellar mass for the various subsamples. This is to emphasize that not all galaxy samples are equivalent in terms of their availability of fuel for star formation, and that this is the case often because of explicit and implicit selection effects, which are in turn also unavoidable in observational studies.

In the lower left panel of Fig.~\ref{fig:sfr}, gas fraction denotes the mass ratio between mass of gravitationally bound gas and stellar mass enclosed within twice the stellar half-mass radius, i.e. $\Mallgravgas/\MSTAR$. For galaxies without a resolvable gas mass, we assign a gas fraction of $10^{-5.0}$, which is the lowest measured gas fraction for any of the studied galaxies. We again indicate the 16th to 84th percentile for each bin as shaded areas for jellyfish galaxies and the `All galaxies' sample. 

The horizontal black, orange and yellow lines mark the minimum gas fraction we impose for selecting the satellites among which to identify jellyfish galaxies; it applies only to the `Inspected satellites', `Jellyfish' and their analogue control samples, but not to `All satellites' or `All galaxies'. As already remarked, it would make no sense to search for jellyfish galaxies, i.e. galaxies exhibiting asymmetric gas tails, among satellites that have already lost all their gas. While we have imposed this selection a priori in our analysis, it is apparent to us that observationally-detected jellyfish also tend towards higher gas fractions compared to the typical satellite population.

All this is reflected in the lower left panel of Fig. \ref{fig:sfr}: there is a clear separation between the general galaxy population (`All galaxies', grey) with a median gas fraction of unity or higher on the one hand, and `Satellite' and `Jellyfish' galaxies and their analogues on the other in TNG50. In fact, the typical gas fractions of the jellyfish galaxies are nearly identical to those of the satellite galaxies (orange vs. red: except for the lowest mass galaxies) and the two analogue samples (by design). Importantly, the `Inspected satellites' are located in between the `All galaxies' and the jellyfish samples, on average. This shows that, even though jellyfish are identified among a certain subsample of satellites, the fact that they undergo RPS does imply that they have lower amounts of gas than their non-jellyfish counterparts. As gas is the fuel for star formation, this also explains why TNG50 jellyfish exhibit typically lower levels of SF than the pool of satellites out of which they have been identified, as we have noted above and in the top panel of Fig.~\ref{fig:sfr}.

\subsubsection{Quenched fractions}\label{sec:quenchedfrac}

The quantification that we have provided above (and that has been used extensively in the jellyfish literature) of the median or average SFRs of jellyfish galaxies in comparison to other galaxies betrays an underlying complexity: because of strong environmental effects, satellite galaxies have often such low levels of star formation that these cannot be measured (or resolved in simulations). This in turn makes the SFR-stellar mass planes bimodal and their interpretation through averages and medians inconclusive. This is why it is typical to assess environmental (and secular quenching) processes by measuring the quenched fractions of galaxies \citep[see e.g.][and all subsequent studies]{Kauffmann2003}. We do so in the bottom middle panel of Fig.~\ref{fig:sfr}.

We measure the fraction of quenched galaxies in bins of stellar mass for the different samples, whereby we count a galaxy as quenched if its sSFR is below $10^{-11}$ yr$^{-1}$, as is common practice in the literature.

As expected, the general population of `Satellite' galaxies (red curve) exhibits the highest quenched fractions, between 0.33 and 0.60, over the whole mass range. In fact, more than half of all $10^{8-9}\,\MSUN$ satellites are quenched. This is expected as galaxies in this sample orbit in the high-density environments of TNG50 groups and clusters and hence undergo the notable environmental effects mentioned in Section~\ref{sec:intro} \citep[see also][for the extensive characterization of the quenched fractions in the IllustrisTNG simulations across galaxy and host mass ranges and across cosmic epochs]{Donnari2019, Donnari2020a, Donnari2020, Joshi2021}.

Crucially, the quenched fractions of jellyfish galaxies, even though they are severely stripped, are strictly lower than that of satellite galaxies, being quenched only in $14-32$ per cent of the cases. This, again, is because of the exclusion of galaxies without any gas, which are instead included in the satellite sample. However, for all mass bins, jellyfish galaxies are more frequently quenched than their `Field analogues' and, at $\MSTAR\lesssim10^{10}\,\MSUN$, than the overall `Inspected satellites'. At $\MSTAR$ between $10^{8.5}$ and $10^{9.5}\,\MSUN$ jellyfish galaxies are also more frequently quenched than their 'satellite analogues'.
Severe RPS therefore seems to increase the likelihood of a galaxy to be quenched compared to its non-stripped counterparts, even though it has no significant effect on the median SFR of those galaxies and even if the gas fractions of jellyfish and their analogues are, by construction, similar. 

Finally, the increase in quenched fractions among the general galaxy population (grey) with increasing stellar mass $\gtrsim 10^{10}\, \MSUN$ is not only consistent qualitatively with observations, but known to be driven by the quenching effects of SMBH feedback in IllustrisTNG \citep{Donnari2020}. This is also the case for the massive satellites of the `Inspected' sample.

\subsubsection{SFR as a function of gas fraction}
We close this analysis by asking: given their gas content, is there any indication that SF proceeds quantitatively differently in TNG50 jellyfish galaxies compared to non-jellyfish galaxies? We address this in the lower right panel of Fig.~\ref{fig:sfr}, where we show the median SFR for the different galaxy samples in bins of gas fraction, in addition to the SFR and gas fraction values of individual jellyfish galaxies as orange dots. Annotations and quantities are as in the previously-discussed panels.

First, the figure shows a general trend of increasing SFR with increasing gas fraction in all galaxies, as is to be expected since higher gas fractions indicate higher available fuel for star formation. Crucially, the correlation between SFR and gas fraction is very similar for all TNG50 galaxies, whether centrals or satellites and jellyfish or not, particularly in terms of normalization. So long as there is some gas (i.e. for gas fractions $>1$ per cent), the different galaxy samples behave similarly to one another and all within their galaxy-to-galaxy variation.  

In summary, the results of the left and right bottom panels of Fig.~\ref{fig:sfr} indicate that the reason we find suppressed SFRs in the case of jellyfish galaxies and satellite galaxies, but not (or at least to a much lower degree) in the case of the `Inspected satellites' sample, is largely due to the lower typical gas fractions in the first two samples. We discuss the implications of these differences in more detail in Section~\ref{sec:discussion}.

\subsection{SFRs across cosmic epochs} \label{sec:SFRredshift}
\begin{figure*}
\includegraphics[width=0.8\textwidth]{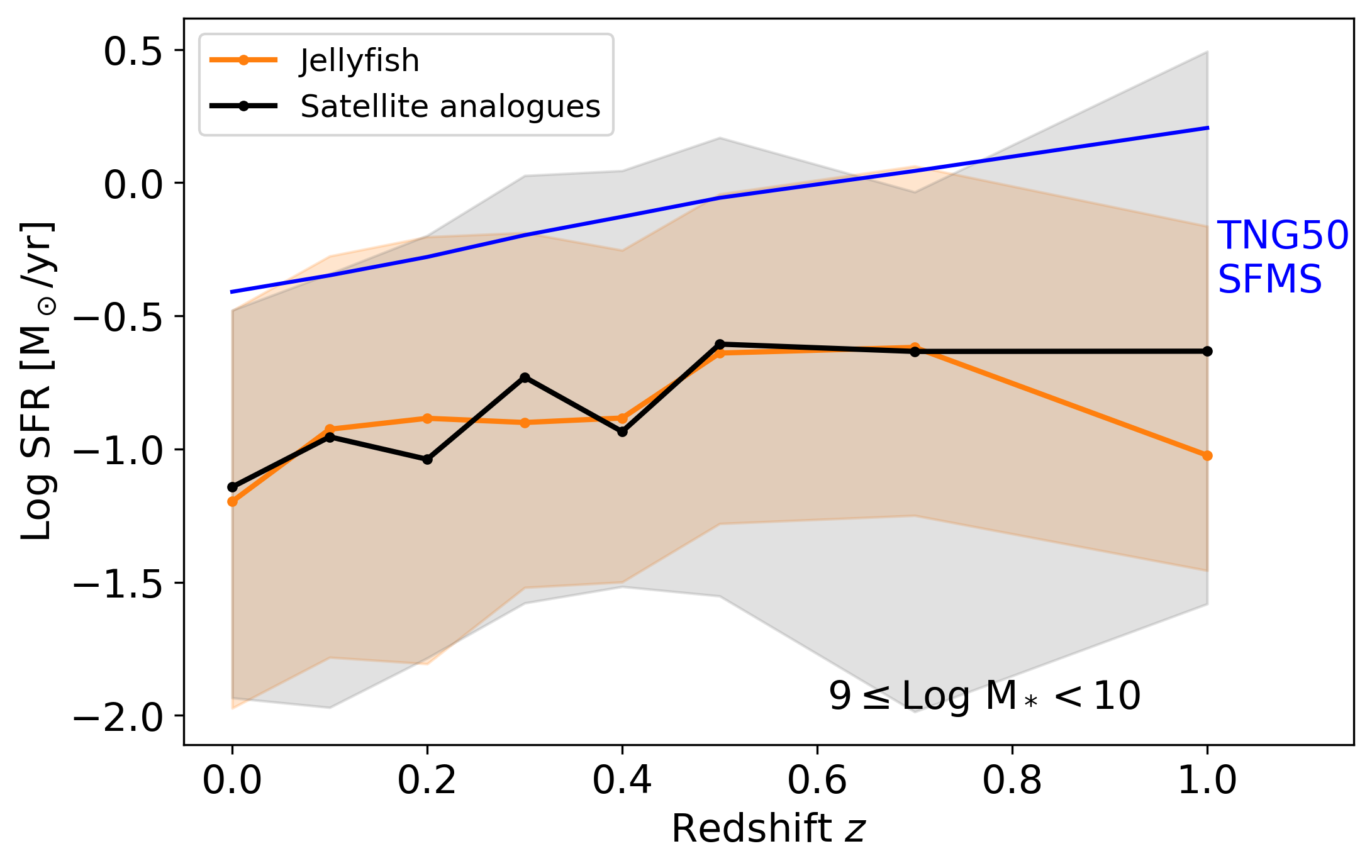}
\caption[SFR as a function of redshift for jellyfish galaxies (orange) and satellite galaxies (black).]{{\bf Star formation activity of populations of TNG50 jellyfish galaxies across cosmic epochs.} We show the median SFR as a function of redshift in `Jellyfish' (orange) and `Satellite analogues' (black) for galaxies with $9\leq \log{(\MSTAR/\MSUN)} <10$. The SFR predicted by the SFMS in TNG50 for this mass bin is denoted in blue. No significant difference between the SFR of jellyfish and satellite analogues can be seen at any redshift.}
\label{fig:SFRredshiftscomp}
\end{figure*}

The TNG50-based results uncovered so far are based on the low-redshift galaxy populations: z = 0, 0.1 and 0.2.
We now consider whether the results of Section~\ref{sec:global_sfr} also hold at higher redshifts, or whether the median SFRs evolve differently compared to the SFMS. 

In Fig.~\ref{fig:SFRredshiftscomp}, we show the median SFR for the jellyfish galaxies and satellite analogues as a function of redshift, for galaxies with mass $\MSTAR=10^{9-10}\MSUN$. For reference, we show the SFR of galaxies on the TNG50 SFMS at each redshift for the same mass bin in blue.

There are two key conclusions to be drawn from Fig.~\ref{fig:SFRredshiftscomp}. First, the median SFR (at constant stellar mass) for jellyfish galaxies does not evolve significantly over time, up to $z \sim 0.4$, after which there is a mild increase at higher redshifts. Note that the results at high redshifts are more susceptible to low-number statistics, due to fewer numbers of massive enough galaxies at these epochs. This lack of evolution of the typical SF activity of jellyfish galaxies at low ($\lesssim 0.4$) redshifts contrasts with the evolution of the SFMS, i.e. the SF activity of the general population of star-forming galaxies, which instead exhibit at least a mildly-increasing SFR with redshift. Second, according to TNG50, there are no significant differences between the jellyfish galaxies and the (stellar mass-, host mass- and gas-fraction-matched) satellite analogues samples at any redshift ($z\sim1$). These results show that the jellyfish galaxies do not show any enhancement in SFR at the population level at any time (since $z=1$ in our study); in fact the overall suppression of their SFRs found at late times in Section~\ref{sec:global_sfr} is seen to be somewhat higher at earlier times, at least out to $z \sim 0.4$.

\subsection{Bursts of star formation in the jellyfish' past} \label{sec:SFRbursts}
\begin{figure*}
\begin{subfigure}[c]{0.4\textwidth}
\includegraphics[width=\textwidth]{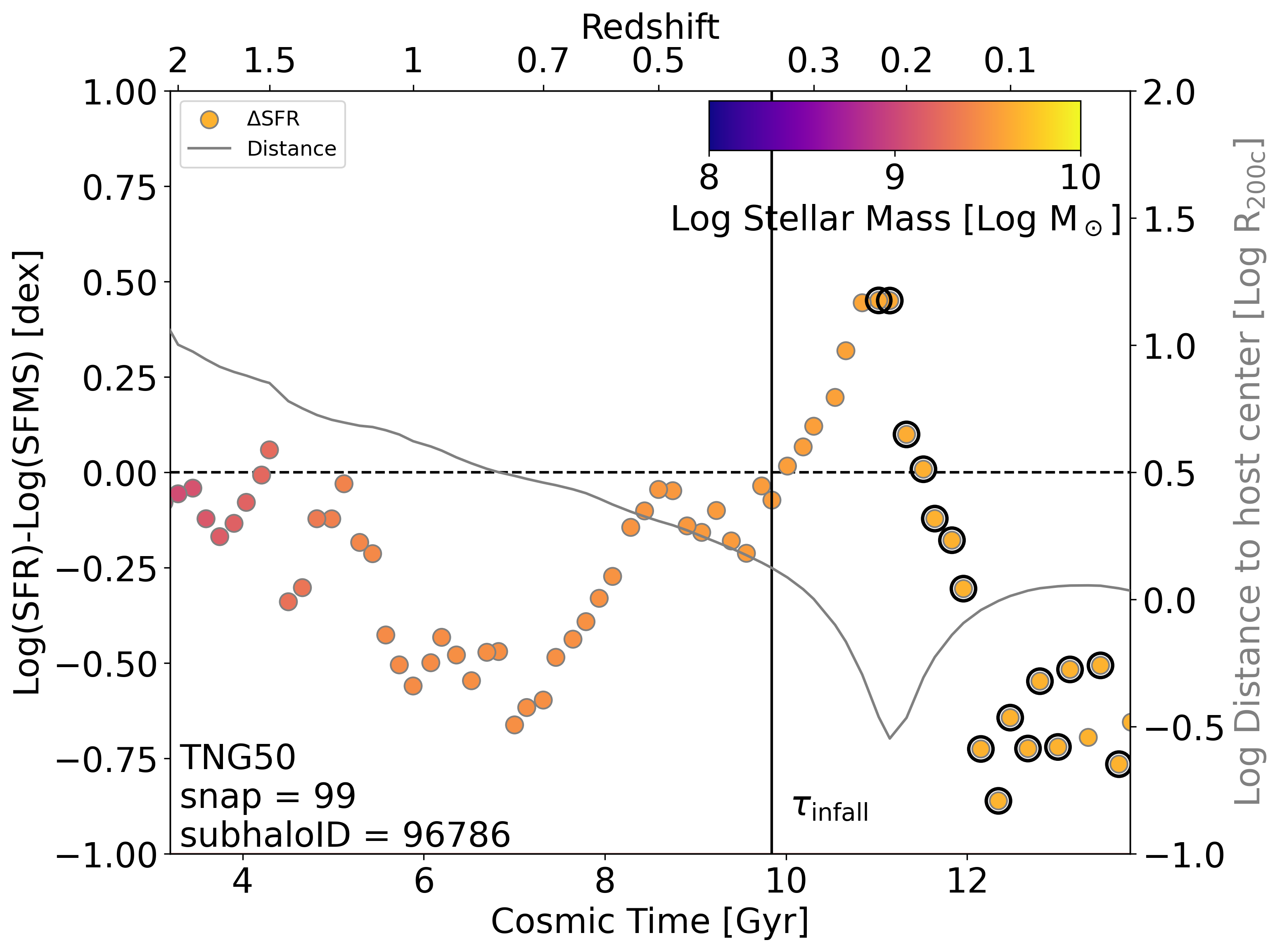}
\end{subfigure}
\hspace{1cm}
\begin{subfigure}[c]{0.4\textwidth}
\includegraphics[width=\textwidth]{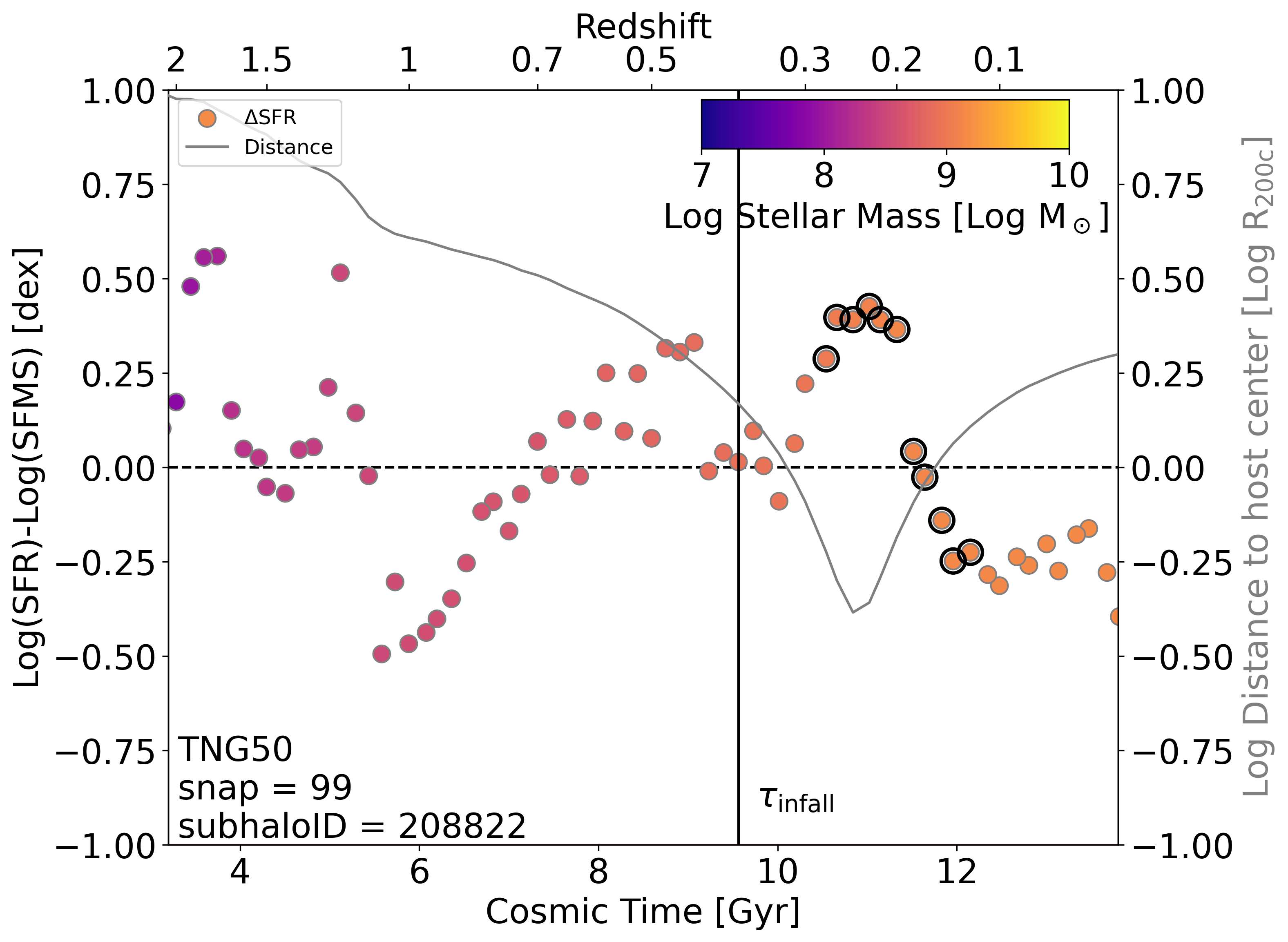}
\end{subfigure}\\
\begin{subfigure}[c]{0.4\textwidth}
\includegraphics[width=\textwidth]{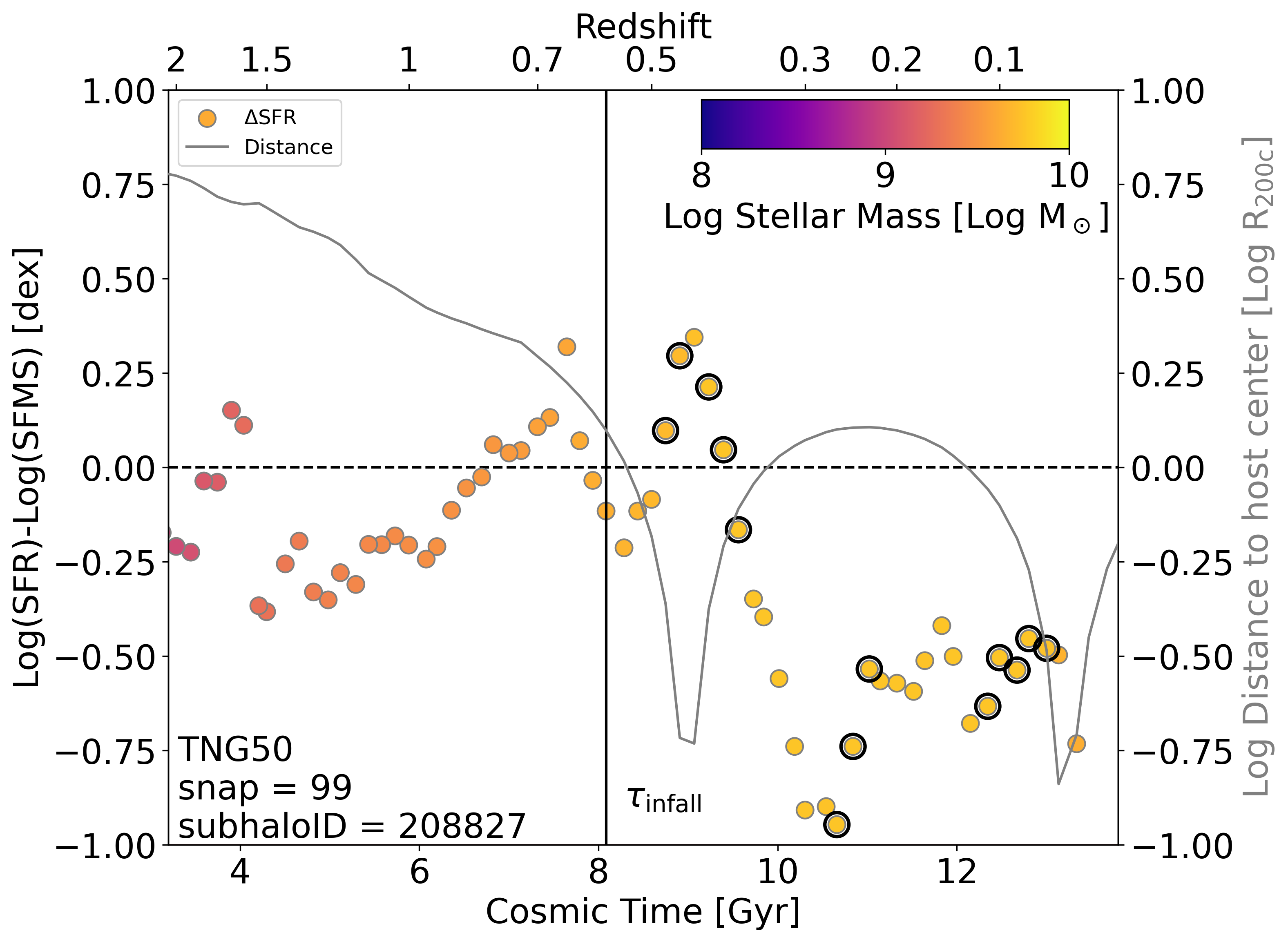}
\end{subfigure}
\hspace{1cm}
\begin{subfigure}[c]{0.4\textwidth}
\includegraphics[width=\textwidth]{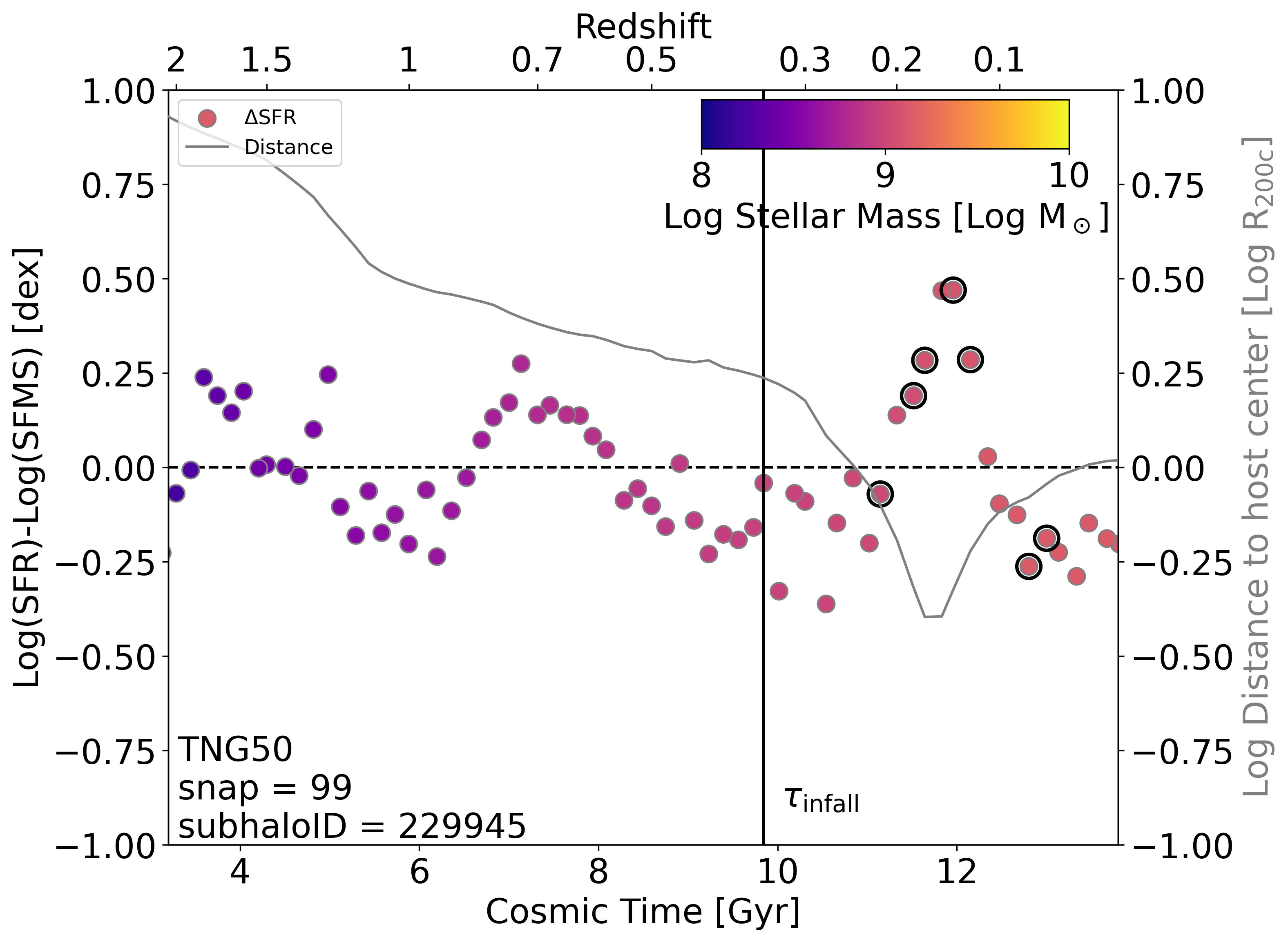}
\end{subfigure}\\
\begin{subfigure}[c]{0.4\textwidth}
\includegraphics[width=\textwidth]{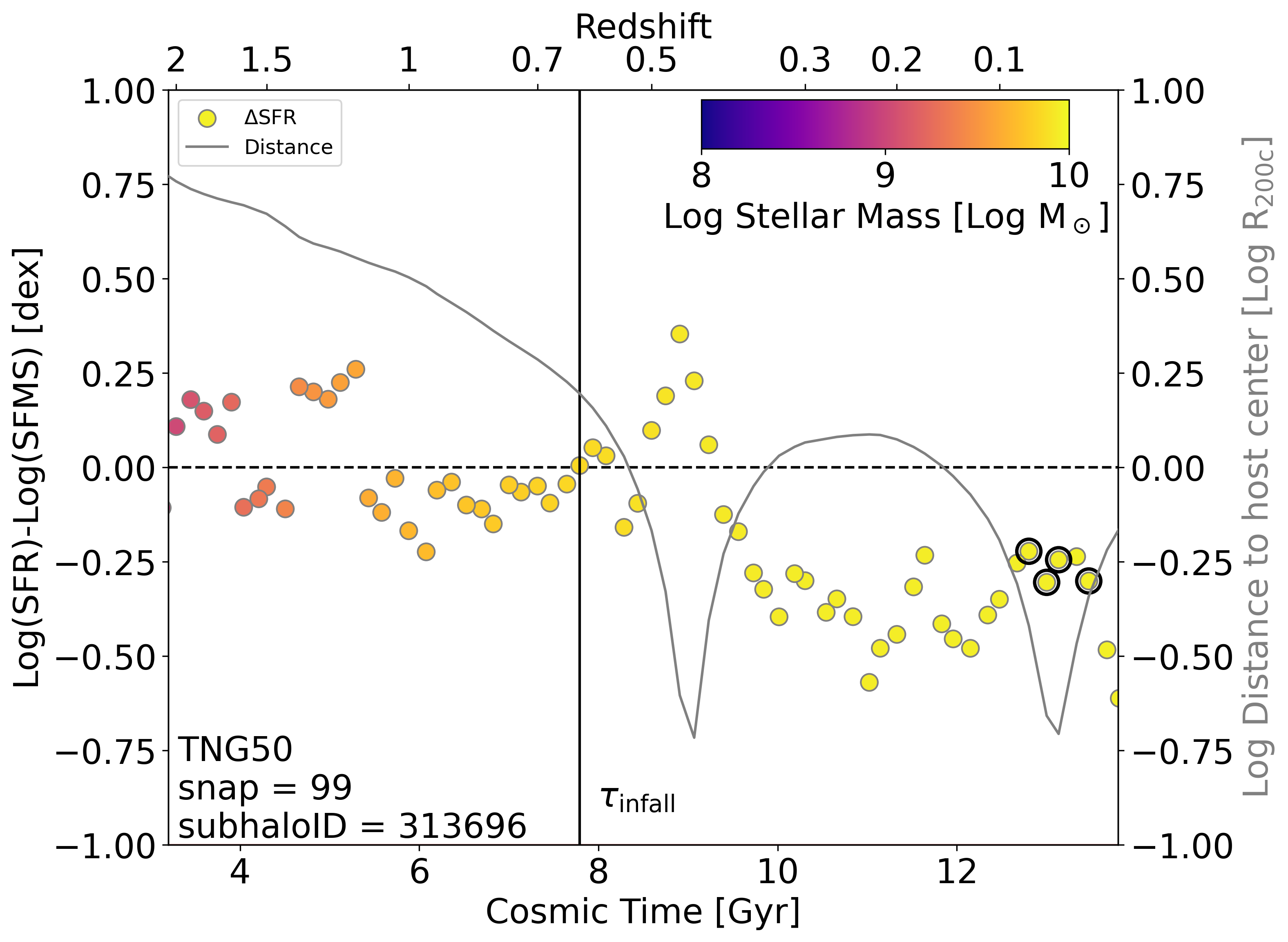}
\end{subfigure}
\hspace{1cm}
\begin{subfigure}[c]{0.4\textwidth}
\includegraphics[width=\textwidth]{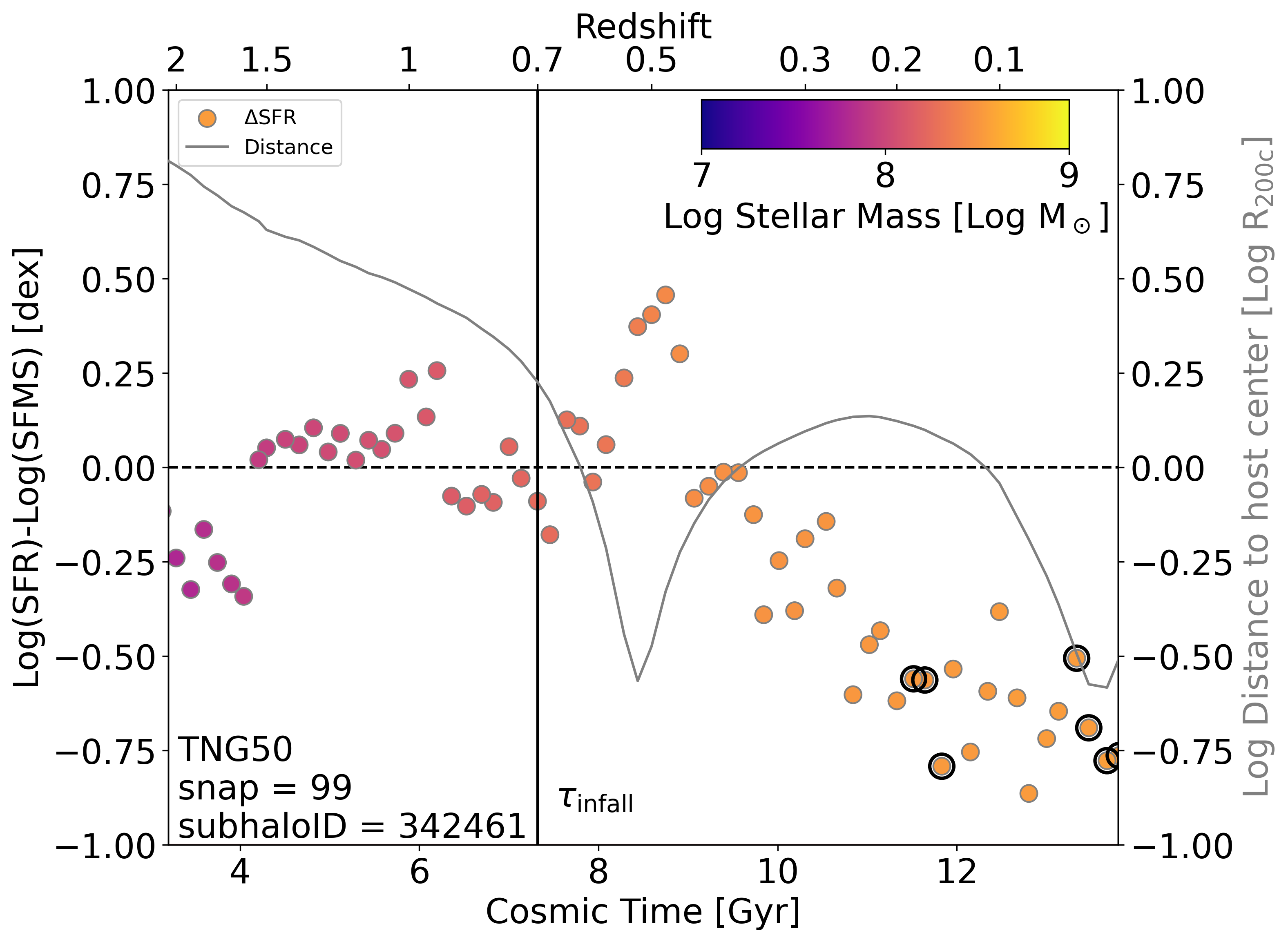}
\end{subfigure}\\
\begin{subfigure}[c]{0.4\textwidth}
\includegraphics[width=\textwidth]{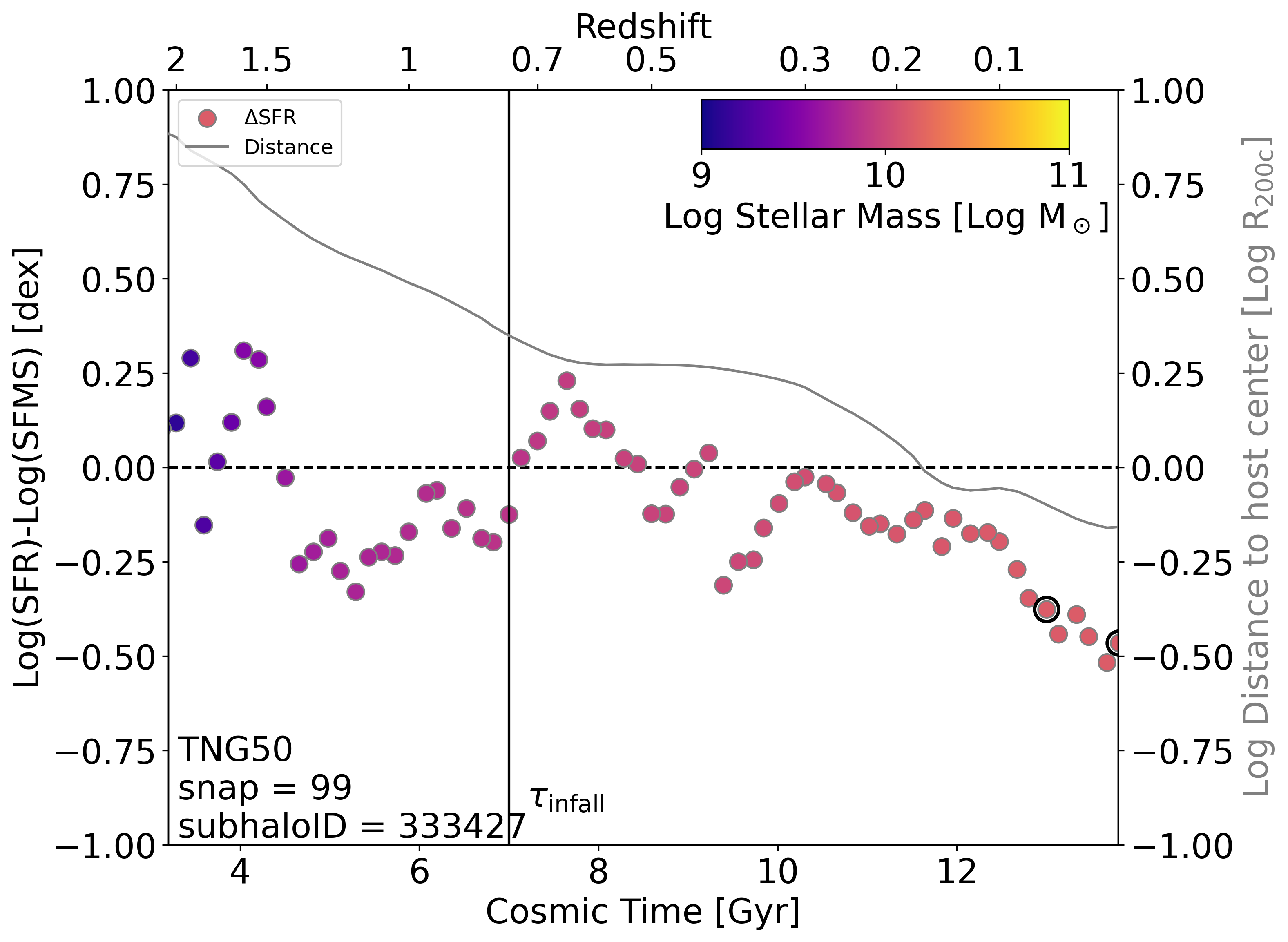}
\end{subfigure}
\hspace{1cm}
\begin{subfigure}[c]{0.4\textwidth}
\includegraphics[width=\textwidth]{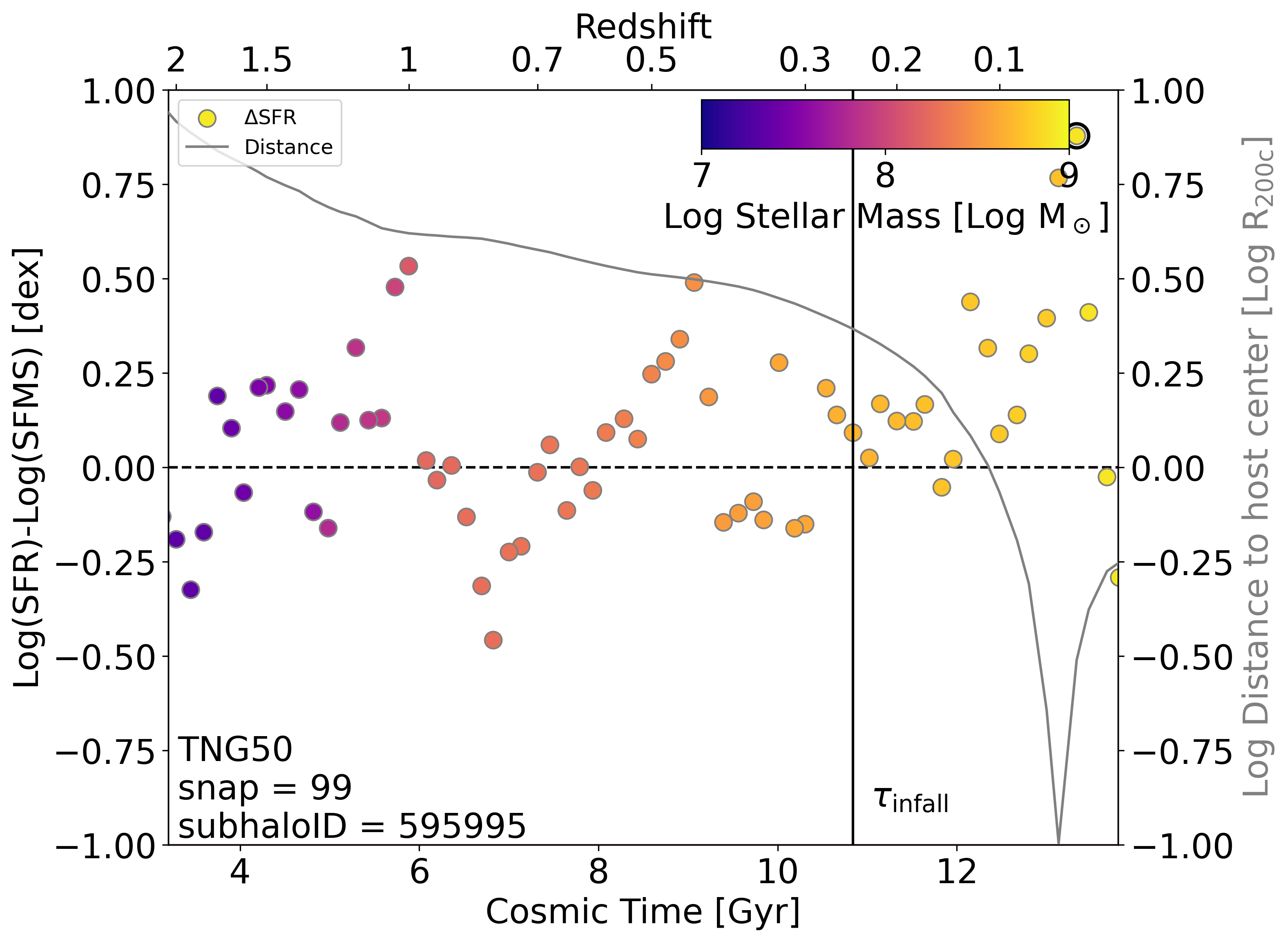}
\end{subfigure}
\caption{{\bf Star formation activity of individual TNG50 jellyfish galaxies across cosmic epochs, i.e. across their individual evolutionary tracks.} We show the offset of SFR from the SFMS ($\Delta$ SFR; colorcoded dots) and distance to the host centre (grey curve) of selected jellyfish galaxies over time. The colour indicates the stellar mass of the galaxy; black circles denote snapshots at which the galaxies were identified as jellyfish galaxies. The vertical line denotes the time of infall of the galaxy into its host halo. Many TNG50 galaxies show peaks in their SFR at some point in their evolution, at which the SFR lies above the SFMS. This peak almost always coincides with the first pericentric passage of the galaxy within the host.}
\label{fig:SFRindividualovertime}
\end{figure*}

Although we do not find an overall population-wide enhanced SFR in TNG50 jellyfish galaxies at late times, TNG50 does predict at least a few galaxies that do have SFRs above the main sequence. Furthermore, the previous results do not exclude the possibility of jellyfish galaxies having temporary bursts of star formation along their evolutionary history. We therefore turn from assessing galaxy populations to studying the evolution in time of individual galaxies. Namely, we inspect the SFRs of individual galaxies over the course of the simulation i.e. along their evolutionary tracks identified by the merger trees (see Section~\ref{sec:tracking} and \citealt{JelRohr} for more details) and compare it to the SFMS at the respective time and mass bin. We have inspected these comparisons for the full sample of jellyfish galaxies. In Fig.~\ref{fig:SFRindividualovertime}, we show a few examples of jellyfish galaxies that demonstrate the various scenarios they encounter.

In each panel, we show the distance of the galaxy's SFR from the SFMS at the given redshift within the mass bin the galaxy is in at the time ($\Delta$SFR; coloured points). The colours indicate the galaxy's stellar mass at that time. Snapshots where the galaxy was identified as a jellyfish galaxy are indicated with additional black circles. Additionally, we also show the cluster-centric distance of the galaxy to its host, normalized by the cluster's virial radius (grey curves). The time at which the galaxy becomes part of its host FoF is indicated by $\tau_{\rm infall}$.

Indeed, we find several galaxies in the TNG50 simulation that are identified as jellyfish galaxies at some snapshot and that experience a sharp increase and decrease, i.e. a peak or burst, of SFR at some point in their lifetime. During this period of increased star formation activity, the SFR can even exceed the SFMS by up to one order of magnitude, even though over large parts of their evolution the galaxies follow the SFMS quite closely. Examples of this phenomenon are shown in the first six panels (rows 1-3) of Fig.~\ref{fig:SFRindividualovertime}. The maximum measured positive difference to the SFMS in any jellyfish galaxy from TNG50 is $1.76\,$dex.
Such short periods of super-SFMS star formation activity can even happen in galaxies that temporarily or overall have a SFR much lower than the SFMS, as can be seen in the two upper left panels of Fig.~\ref{fig:SFRindividualovertime}.

Comparing the evolution of the $\Delta$SFR to the galaxies' distance from the host centre, it is apparent that, at least after infall, peaks in SFR almost always correlate with the closest proximity of the galaxy with the host center, i.e. during pericentric passages. However, usually only the first approach to the host leads to a super-SFMS increase in SFR, while the subsequent ones do not result again in an equally increased SFR. Examples for this are shown in Fig.~\ref{fig:SFRindividualovertime}, panels 3, 5 and 6. We speculate that this peak in SFR is likely caused by gas compression \citep{Compression2, Compression1}, which disrupts the pressure balance stabilizing the gas against collapse. These effects may get increasingly stronger as the galaxy traverses regions of higher ambient-gas density and approaches the central galaxy, hence peaking at the pericentric passage. During subsequent pericentric passages, a second or third peak in star formation may most probably be suppressed because the galactic gas has already been used up during the first star formation burst or stripped during the orbit around the central galaxy.

Within our full sample of TNG50 jellyfish galaxies, the vast majority experiences an epoch of increased SFR during their lifetime. We have inspected all evolutionary tracks of TNG50 galaxies that have been flagged as jellyfish at some point in time and we find that about 74 per cent have a $\Delta$SFR of $\geq 0.5$~dex at some point during their evolution and about 30 per cent of the jellyfish galaxies experience such an increased SFR at $z\leq 2$. 

It is noteworthy that most jellyfish galaxies are identified as such in snapshots following these peaks in $\Delta$SFR during a pericentric passage, or at least during its descending flank i.e. during the phase of decreasing SFR. The upper two panels of Fig.~\ref{fig:SFRindividualovertime} show examples of such behaviour, which we find to be typical for the overall sample. In fact, of the galaxies that experience a $\Delta$SFR increase of $0.5\,$dex or higher at $z\leq 2$, (upon visual inspection) about 70 per cent are identified as jellyfish galaxies close to their time of a pericentric passage. Those identifications then usually fall within an epoch of decreasing SFR in the galaxy. However, one has to keep in mind that even though many galaxies in the sample show the aforementioned behaviour, there are still a number of jellyfish galaxies that show a SFR history with low SFR variability and without peaks. An example of this is shown in the bottom left panel of Fig.~\ref{fig:SFRindividualovertime}. 

Overall, we find that although the jellyfish galaxies do not show enhanced global SFRs at late times at the population level, the majority of individual jellyfish galaxies do exhibit short periods of SFRs above the SFMS i.e. starbursts, usually during pericentric passages. Additionally, it appears that in most cases, the galaxies are identified as jellyfish galaxies during the decreasing leg of such a burst, or soon after it. The former scenario then implies that within the TNG50 model, there are indeed some jellyfish galaxies that have enhanced global SFRs. An in-depth analysis of the evolution of individual jellyfish galaxies, and specifically, of when and where RPS occurs, is quantified and discussed by \citealt{JelRohr}.

\section{Discussion} \label{sec:discussion}

\subsection{Do jellyfish produce stars at higher rates than other satellites and field galaxies, according to TNG50?} \label{sec:discussion_observations}

The main aim of this paper is to investigate whether galaxies undergoing RPS, i.e. jellyfish galaxies, within TNG50 groups and clusters and over a wide range of stellar masses, have increased SFRs, as seen in several (but not all) observational studies of jellyfish galaxies. From the analysis of the global, galaxy-wide SFRs of jellyfish galaxies and different control samples in Section~\ref{sec:global_sfr}, we in fact do not recover in TNG50 the population-wide increased SFR of jellyfish galaxies compared to satellite and field galaxies found in observations by e.g. \citealt{Vulcani2018}, \citealt{Ramatsoku2020} and \citealt{Vulcani2020}. Instead, TNG50 jellyfish galaxies typically show a global SFR comparable to their control samples at a given stellar mass, and lower by $0.6-0.8$~dex at $\MSTAR\lesssim10^{10}\,\MSUN$ compared to the general star forming galaxy population in the simulation ($0.3-0.7$~dex compared to the sample of `Inspected satellites' from which they have been identified). This remains true for all redshifts studied in this work, i.e. $z\lesssim1$ (Section~\ref{sec:SFRredshift}). The SFRs are similarly lower than the SFMS for the jellyfish, all satellites and control samples at fixed gas fractions, specifically and typically for galaxies with stellar masses of $10^{9-10}\,\MSUN$. At higher masses, TNG50 jellyfish typically behave as the general galaxy population. The outcomes of TNG50 seem to be more similar to the findings of \citealt{Yoon2017} and \citealt{Mun2021} or the outcome of the study by \citealt{Roberts2021a} in galaxy groups.

We believe that this apparent contention between our TNG50-based results and those of the GASP survey are at least partly driven by selection effects and the precise methods employed to identify jellyfish galaxies. Additionally, it may be driven by the different quantities being considered. For example, \citealt{Vulcani2020} find increased SFRs when comparing only resolved \emph{star-forming} regions between the jellyfish and control samples;  \citealt{Ramatsoku2020} find enhanced SFRs for jellyfish galaxies at fixed \emph{HI gas mass}; on the other hand, \citealt{Moretti2020} have shown that, when the total gas mass is considered, jellyfish galaxies have SFRs in line with other star-forming galaxies, in agreement with our results in Fig.~\ref{fig:sfr}. 
Furthermore, the GASP survey targets satellites in clusters of mass $\MHOST=10^{13.6-15.2}\MSUN$ which is markedly higher than in our sample, and initially identifies gas-stripping candidates as galaxies that have signs of debris/morphological disturbances based on B-band imaging \citep{Poggianti2017}. Furthermore, the control sample consists of galaxies that have no such optical signs of gas stripping. The survey then obtains complete IFU data for the candidate galaxies with MUSE, including the tail regions, thus enabling the authors to analyse the properties of the stellar and gaseous components in detail. In contrast, in our study, we include all galaxies that have a minimum gas content in our `Inspected satellites' sample and jellyfish galaxies are identified based on the gas mass column density maps (relative to contours of stellar density maps), including all the gas, irrespective of phase and without imposing any surface density or brightness detection limit. It is therefore possible that the GASP candidate selection does not capture all jellyfish galaxies, i.e. galaxies that may have smaller gas tails but do not show obvious signs of morphological disturbances in optical imaging, although the extent to which this may affect the results is unknown. However, it is at least possible that this biases the GASP jellyfish galaxies towards more extreme stripping conditions. The SFR enhancements found by \citealt{Vulcani2018,Vulcani2020} are expected to be caused by gas compression along the leading edge of the jellyfish galaxies; whether this process occurs preferentially in more extreme stripping conditions remains to be seen. Finally, we notice that, by comparing the maps of Figs.~\ref{fig:maps_column} and \ref{fig:maps_sfr} and without including any observational realism and limitations, it is clearly easier according to TNG50 to identify jellyfish galaxies based on gas phases that do not trace star formation and very dense gas than e.g. H$\alpha$.

Furthermore, the outcomes of comparisons like the one done in Section~\ref{sec:global_sfr}, are highly sensitive to the definition of the control sample and any additional cuts applied to the data. Changes in the selection criteria of the control samples are able to strongly alter the outcome of any SF-activity comparisons between jellyfish galaxies and the control sample. We have shown this complexity by comparing galaxy samples with different gas fractions i.e. with different availability of fuel for star formation: implicit or explicit selection biases may affect the gas fraction distributions in galaxy samples and thus may imply higher or lower SF activity in populations of jellyfish galaxies even if their jellyfish nature (or the fact that they undergo RPS) may not be the physical cause per se of a different mode of star formation.

On the other hand, TNG50 does predict a fraction of jellyfish at any given time to be above the star forming main sequence. It also predicts bursts of star formation during the evolution of about three quarters of all jellyfish galaxies (Section~\ref{sec:SFRbursts}). Jellyfish galaxies in general are rather rare objects; however, we speculate that, depending on the detection method, star bursting galaxies may be more prominent in observations, thus resulting in a similar selection bias as mentioned above, whereby observed jellyfish galaxies are found to have enhanced SFRs. That galaxies could be missing from observations is also implied by the larger relative number of jellyfish galaxies found in IllustrisTNG compared to observations. In \citealt{JelZinger}, we find a jellyfish fraction of about 8 per cent in TNG50 among the inspected satellites; on the other hand, e.g. \citealt{Poggianti2016} find a total of 2 per cent of galaxies from their observational sample from 76 galaxy clusters and 176 galaxy groups to be jellyfish candidates. Even if these fractions strongly depend on the parent sample of satellites among which the search for jellyfish is made, the jellyfish frequency uncovered with the IllustrisTNG simulation is completely agnostic to gas phases and does not account for observational effects and limitations. A preferential observation of star-bursting galaxies could explain the deviation between results from simulations and observational results.

\subsection{Possible limitations of the current work and looking ahead}

TNG50 has provided us with a sample of jellyfish galaxies unmatched in size by any observational survey and theoretical analysis, which in turn has allowed us to conduct a statistically robust investigation into the star-formation properties of jellyfish and non-jellyfish galaxies. However, we caution that some of the results presented above are driven by the specifics of the IllustrisTNG model itself, as well as possible other limitations. We discuss these in detail here, although we note that most of these issues affect the SFRs in the tails of the jellyfish galaxies, and have less of an impact on the global SFRs and the associated comparisons. Furthermore, despite the over-simplicity of the ISM and star formation model underlying TNG50, the IllustrisTNG model has been shown to return galaxy star formation activities and satellite quenched fractions in the ball park of observations in the regimes studied here \citep[e.g.][]{Donnari2019, Donnari2020, Donnari2020a, Joshi2021}, thus lending credibility to the effective outcome of the model when assessed across galaxy populations (see also \citealt{Kukstas2022} for the comparison of environmental effects across different simulations).

\subsubsection{Star-formation and ISM model in IllustrisTNG}

As mentioned in Section~\ref{sec:tng50}, the IllustrisTNG simulations, including TNG50, are based on a simplistic non-multi-phase modeling of the ISM. Furthermore, the process of star formation in the IllustrisTNG model is stochastic in nature, i.e. if the density in a gas cell rises to exceed a chosen minimum density threshold, a star particle is formed with a certain probability \citep[see][]{SpringelHern2003, Pillepich2018}: it is designed to follow a Kennicutt-Schmidt relation under the assumption of a Chabrier Initial Mass Function \citep{Chabrier2003b, Chabrier2003a}. Other gas properties such as temperature, magnetic field, pressure, the presence of flows or gravitational potentials, which might hinder or support the formation of stars, are not explicitly included in the recipe for star formation, i.e. not below the spatial resolution of the star-forming gas cells; its stochasticity, the associated probability, and the chosen density threshold for star formation are designed to account for these other factors, in conjunction with stellar feedback. However, the star-formation process is ultimately (approximately) calibrated to reproduce certain observed galaxy scaling relations at $z=0$, especially for isolated galaxies, and are therefore most appropriate for star formation within the main body of a galaxy. Such recipes do not necessarily account for the unique conditions encountered in the tails of the jellyfish galaxies in group and cluster environments, which may affect the SFRs we recover in the jellyfish galaxy tails. However, this should not impact the overall results regarding the global i.e. galaxy-wide SFRs of the jellyfish galaxies, which as we and others have found, are still dominated by the main body of the galaxy. 

\subsubsection{Spatial resolution effects}
A defining characteristic of a quasi-Lagrangian moving-mesh hydrodynamical code like \textsc{arepo}, which is the underlying framework of IllustrisTNG, is that the sizes of the gas cells in the simulations depend on the local density of the gas \citep[see e.g.][]{Pillepich2019, Pillepich2021}. While this allows for increased resolution in the dense central regions of galaxies, it also means that regions of low density are also less spatially resolved than higher density regions. This may be particularly important in the tails of the jellyfish galaxies, which are likely to exhibit lower gas density than the galaxies' bodies and to hence be realized by larger gas cells. If the physical conditions captured by the IllustrisTNG galaxy formation model could lead to the formation of small knots of dense gas within the tails that may be star forming (as the H$\alpha$ knots observed in real tails, e.g. by \citealt{Vulcani2018} and \citealt{Poggianti2017}), then the \textsc{arepo} code would in principle be able to adapt its mesh and to increase the spatial resolution according to the density \citep[as is the case for the cold small gas clouds in the circumgalactic medium of TNG50 galaxies][]{Nelson2020, Ramesh2023b}. However, ultimately the spatial resolution in the TNG50 galaxies is limited and these knots may remain under resolved. These effects in turn may lead to the TNG50 SFRs in the tails being underestimated, although again, the impact on the global SFRs is expected to be minimal.

\subsubsection{Temporal resolution effects}
In Section~\ref{sec:SFRbursts}, we explored the star-formation histories of individual jellyfish and showed that the majority of the unique-branch jellyfish galaxies in our sample have experienced bursts of star formation during their histories, usually correlated with their first pericentric passage within the host cluster. This analysis was based on the instantaneous SFR of the galaxy at a given snapshot. Therefore, the analysis is dependent on the time interval between the available consecutive snapshots, which is approximately $\sim 150$ Myr (note that this is separate from the temporal resolution enforced during the simulation run, which is considerably higher). This could in principle prevent us from capturing shorter starbursts that may have occurred between the available snapshots. From Fig.~\ref{fig:SFRindividualovertime} it would appear that the peaks or bursts of star formation have long duration, encompassing multiple snapshots. Yet, the possibility of the existence of SF bursts of shorter duration can only be assessed with higher-cadence outputs or by relying on the stellar particle ages; we postpone to future work the task of assessing this.

\subsubsection{Measurement of SFRs}
Another source of the discrepancies between our results and observations could be the measurement of the SFRs themselves (see also Appendix~\ref{appendix}). In simulations, we are able to directly measure the SFRs as we have access to the full star-formation histories of the galaxies, whereas observations rely on proxies for SFR, such as the strength of H$\alpha$ emission or IR luminosity, which are then calibrated to measure the SFRs of the galaxies, but are sensitive to star formation on different timescales. Therefore, the observed SFRs are dependent on the precise calibration being used and, while these calibrations are usually robust for the overall SFRs of galaxies, they may not be appropriate for the SFRs in unique environments like jellyfish tails. In fact, H$\alpha$ might not be as tightly linked to star formation in jellyfish tails as it is in galactic disks, as found by some studies \citep{Boselli2016, Cramer2019}.

\section{Summary and conclusions} \label{sec:summary}

In this paper we have quantified the star formation activity of jellyfish galaxies realized in a large-scale cosmological magnetohydrodynamical simulation of galaxies: TNG50 \citep{Nelson2019, Pillepich2019}.

In particular, we have focused on an unprecedented sample of 780 jellyfish in selected snapshots at $z\le1$: these were visually identified via our Cosmological Jellyfish Zooniverse project \citep{JelZinger} among more than 50000 satellites with stellar mass $> 10^{8.3}\,\MSUN$, belonging to groups and clusters of $\MHOST = 10^{10.5-14.3}\,\MSUN$, and with a gas mass fraction $>$ 1 per cent.

We contrasted the jellyfish star formation rates (SFRs) against various populations of galaxies in the simulation, including the general galaxy population and its star-forming main sequence (SFMS), the pool of satellite galaxies inspected for identification, all satellite galaxies in the simulation, and two carefully constructed control samples of both satellite and field galaxies that are not jellyfish but are most similar to the jellyfish galaxies in stellar mass, gas fraction and host mass (Sections~\ref{sec:zooniverse}, \ref{sec:samples}, Fig.~\ref{fig:ovsample} and Table~\ref{tab:Jelnumbers}).
We have measured global i.e. galaxy-wide SFRs for each galaxy as well as the SF occurring separately in the main body and in the ram-pressure stripped tails of jellyfish (Section~\ref{sec:disk_tail}). We have compared galaxy samples at the population level (Sections~\ref{sec:global_sfr} and \ref{sec:SFRredshift}) as well as followed individual galaxies across their evolutionary tracks (Section~\ref{sec:SFRbursts}).\\

Our main results are as follows:

\begin{itemize}

\item According to TNG50, despite being severely affected by ram-pressure stripping, jellyfish galaxies are typically star forming (Figs.~\ref{fig:maps_column}, \ref{fig:maps_sfr}, \ref{fig:sfr_bodytail} and \ref{fig:sfr}). This occurs even though our identification of jellyfish galaxies is completely agnostic to tracers of star formation.

\item In fact, some star formation also occurs in the ram pressure-stripped tails, even though in TNG50 this unfolds at much subdominant rates in comparison to that in the main jellyfish bodies (Fig.~\ref{fig:sfr_bodytail}).

\item Within the TNG50 jellyfish samples, it is possible to identify examples at any cosmic epoch whose SFRs exceed the ridge of the SFMS (Figs.~\ref{fig:sfr_bodytail} and \ref{fig:sfr}).

\item However, TNG50 predicts no overall, population-wide enhancement of star formation in jellyfish galaxies in comparison to the population of satellites inspected for identification or field galaxies at the same mass or redshift. Rather, for a given stellar mass, the median SFR of TNG50 jellyfish galaxies is below that of the SFMS and comparable to that of control samples of field and satellite analogue galaxies with similar gas fractions (Fig.~\ref{fig:sfr}, top). These findings hold up to $z\sim1$ (Fig.~\ref{fig:SFRredshiftscomp}).

\item Importantly, TNG50 jellyfish galaxies are less frequently quenched than the general population of satellite galaxies (Fig.~\ref{fig:sfr}, bottom): this naturally follows from the fact that jellyfish galaxies are biased towards larger gas fractions than the randomly-selected satellites (otherwise they could not exhibit tails of stripped gas).

\item Finally, many TNG50 jellyfish galaxies experience phases of enhanced SFRs during their evolution (Fig.~\ref{fig:SFRindividualovertime}). In particular, more than 74 per cent of TNG50 jellyfish have experienced, at some point during their evolutionary paths, short periods of time when their SFRs are above the SFMS, i.e. starbursts, usually during pericentric passages. However, these do not translate into a population-wide enhancement at any given epoch.

\end{itemize}

Our TNG50-based findings are qualitatively in contrast to some, but not all, observational findings on the star-formation activity of jellyfish galaxies. The unmatched sample sizes provided by TNG50 have allowed us to conduct a statistically robust investigation into the star-formation properties of jellyfish and non-jellyfish galaxies: in particular, we have emphasized the importance of accounting for implicit and explicit selection biases that undoubtedly affect any jellyfish sample, whether simulated or observationally identified. In future work, we will go beyond the simplified numerical treatment of the ISM and star formation in TNG50 and will assess the impact of the numerical and physical models in predicting the star formation in jellyfish galaxies, particularly in their tails.

\section*{Acknowledgements}

JG acknowledges financial support from the European Research Council (ERC) via the ERC Synergy Grant "ECOGAL: Understanding our Galactic ecosystem -- From the disk of the Milky Way to the formation sites of stars and planets" (grant 855130). This publication uses data generated via the \href{https://www.zooniverse.org/}{Zooniverse.org} platform, development of which is funded by generous support, including a Global Impact Award from Google, and by a grant from the Alfred P. Sloan Foundation. JG and ER are fellows of the International Max Planck Research School for Astronomy and Cosmic Physics at the University of Heidelberg (IMPRS-HD). GJ acknowledges funding from the European Union’s Horizon 2020 research and innovation programme under grant agreement No. 818085 GMGalaxies.

The TNG50 simulation was run with compute time granted by the Gauss
Centre for Supercomputing (GCS) under Large-Scale Projects GCS-DWAR on the
GCS share of the supercomputer Hazel Hen at the High Performance Computing
Center Stuttgart (HLRS).

\section*{Data Availability}

All data of the TNG simulations, including TNG50, are publicly available at \url{https://www.tng-project.org/} and described in \citet{Nelson2019}. The scores for the identification of jellyfish galaxies are also publicly available \citep{JelZinger}.\\



\bibliographystyle{mnras}
\bibliography{references} 



\appendix

\section{Different measures for SFR} \label{appendix}

\begin{figure*}
\includegraphics[width=0.8\textwidth]{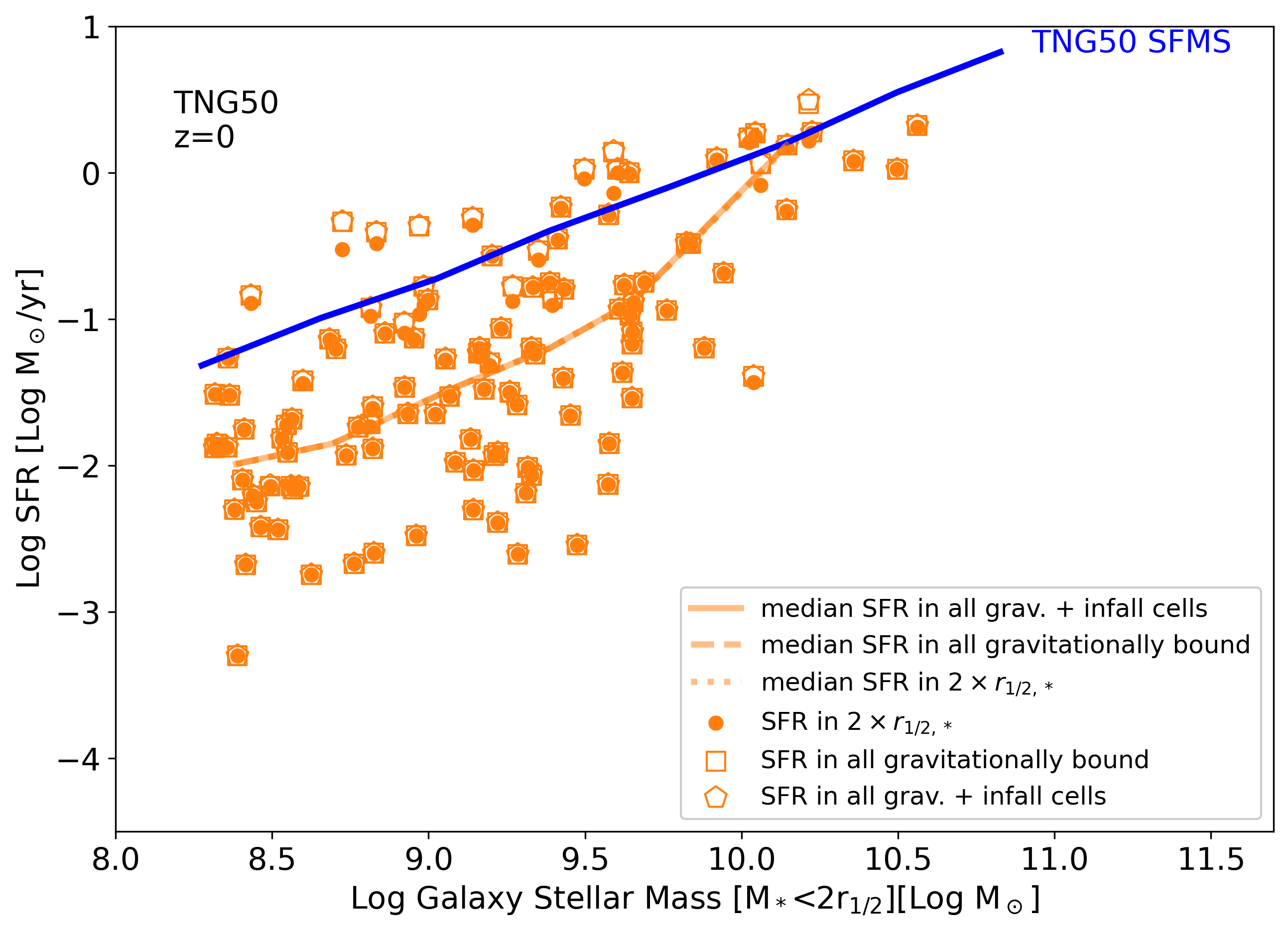}
\caption{ Global SFR as a function of galactic stellar mass for jellyfish galaxies showing SFR within 2$\times$R$^*_{1/2}$ (dot, dotted line), all gravitationally bound gas (square, dashed line) and with additional gas cells from galaxy infall (pentagon, full line). Changes in SFR caused by a change of the galaxy's volume being taken into account for the measurement are barely noticeable. Similarly, taking additional cells from the time of infall into account does not result in any noteworthy change in SFR.}
\label{SFRdiffmes}
\end{figure*}

In this section, we assess various methods of measuring SFRs within IllustrisTNG to study the robustness of our results in terms of SFR.

\subsection{Influence of Accounted Galaxy Volume}
The IllustrisTNG results already provide several different measurements of SFR, as mentioned in Section~\ref{sec:globalsfr}. The results presented in the paper make use of the SFR within all gravitationally bound gas cells. A second possibility is to restrict the measurement to within $2 \times \rhStar$, which is approximately the region where the stellar component of the galaxy would be observable. Although this latter value is by design lower than the former, we include it to demonstrate that, in fact, this choice does not have any significant impact on our results. 

One final consideration in determining the SFRs, particularly of jellyfish galaxies, is that by only considering the gas cells \emph{currently} gravitationally bound to the galaxy, we may be missing some gas that, although considered detached from the galaxy by \textsc{subfind}, may in fact still be within the tail. We therefore consider a third measurement of the SFR, which includes gas that was previously bound to the galaxy. To do so, we add the SFRs from gas cells that were bound to the galaxy at the time of infall into the host halo to the SFRs from the currently bound gas cells. The precise method of tracking these detached gas cells is described in Section~\ref{sec:InfallCells}.

In Fig.~\ref{SFRdiffmes} we contrast the three different SFRs as a function of stellar mass for redshift $z=0$. As one can see, the different modes of SFR measurement do not greatly alter the overall global SFR. The SFR within $2 \times \rhStar$ (dots) seems to already account for the majority of star formation occurring within the galaxy. The extension of the included volume to that of all gravitationally bound gas (squares) results in at best a mild change in the SFR; in fact the difference is only noticeable for the jellyfish with the highest SFRs ($>10^{-2}\,\MSUN$yr$^{-1}$) and negligible for the majority of the sample.

Similarly, adding the SFR of gas cells which were bound to the galaxies at infall but have since become unbound (pentagons) to that of gravitationally bound cells results in no noticeable change in the SFRs. We therefore conclude that the added gas cells, which mainly contribute to the tails of jellyfish galaxies, do not have a major impact on the global SFR and, by themselves, show low star formation activity.

From these results, we conclude that most of the star formation in the jellyfish galaxies appears to take place within 2$\times$r$_{1/2, *}$. Gravitationally bound gas cells outside of this volume contribute little to the global SFR, and gas cells that are not gravitationally bound add no noteworthy star formation to the galaxy. As we are interested in jellyfish galaxies as a whole, including their tails, we include SF from \emph{all} gravitationally bound gas cells for the measurement of SFR, since relying on the SFR within $2 \times \rhStar$ would lead us to underestimate their SFRs, especially in the case of massive galaxies which have gas components extending as far out or even farther than the stellar component. Adding gas cells from infall draws a more complete picture of the jellyfish tails, but as the presented method (see Section~\ref{sec:InfallCells}) is not very rigorous because of the change of cell IDs in the IllustrisTNG simulations, and the additional cells do not seem to have a significant influence on the global SFR compared to the SFR measured in bound gas (compare squares and pentagons), and also not in the tail (compare dots and pentagons), we refrain from using this addition in the main analysis.

\subsection{Accounting for the gas cells from the time of Infall}\label{sec:InfallCells}

\begin{figure}
    \centering
    \includegraphics[width=0.45\textwidth]{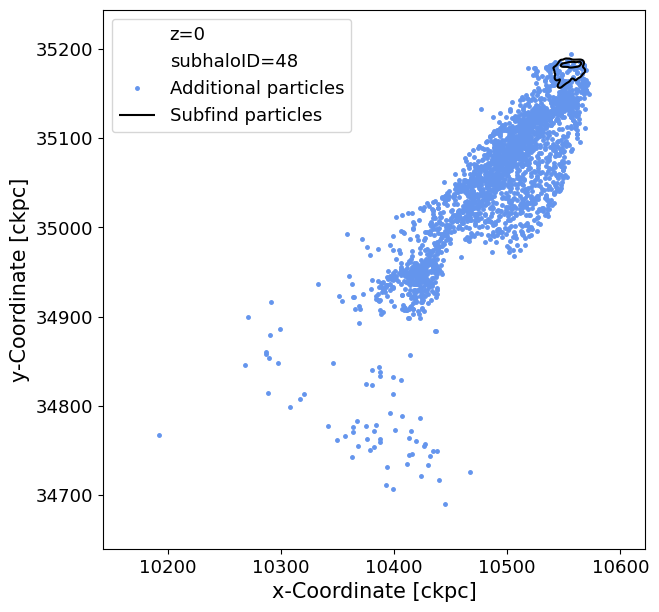}
    \caption[Comparison of gas cells found by Subfind algorithm and additional gas cells from galaxy infall time.]{Galaxy 48 of TNG50 at z=0. Gas cells identified by the Subfind algorithm are given as black contours. Blue dots denote additional cells identified by the search for infall cells. Additional cells range far out from the region of identified Subfind cells.}
    \label{additionalparticles}
\end{figure}

To track cells which were part of the galaxy at its time of infall, we first need to determine the infall time of each galaxy. This has already been done by \citealt{Chua2017}, who created a catalogue containing the desired information, the 'InfallCatalog'. IllustrisTNG provides merger trees that connect galaxies identified by \textsc{subfind} through the "SubLink\_gal" algorithm of \citealt{Rodriguez-Gomez2015}. However, the merger trees do not contain information on the merger histories of the host FOF groups. Instead, to create the InfallCatalog, the central galaxy of each host halo is traced back in time and it is assumed that there is a direct correspondence between the main branch of the host halo and the main branch of its central galaxy. Satellite galaxies are then also traced back in time to the last snapshot in which their progenitors are not in the same FOF group as the progenitors of the central galaxy in their current host halo. 

Thus, we obtain the progenitors of each jellyfish galaxy at its time of infall, query the cell IDs of the gas cells belonging to the galaxy at infall and then trace the same cell IDs to the current snapshot. Note that in doing so, we make the assumption that each gas cell keeps the same cell ID for all times, which is not necessarily the case. However, this procedure can be justified in low density regions with low star-formation activity, where processes changing cell IDs rarely occur.

The result of tracking these initial gas cells for an example galaxy is presented in Fig.~\ref{additionalparticles}. We find that gas cells traced from the time of infall (blue dots) build a low density tail behind the galaxy, which extends to a much larger volume than that covered by the gas cells currently assigned to the galaxy by the Subfind algorithm (black contours). This is the case for many of the galaxies in our jellyfish sample. These additional gas cells could potentially alter several gas properties of the galaxies we study. However, as shown in the previous section, we find that the inclusion of these gas cells has a negligible impact on the SFRs of the galaxies and therefore, do not affect the results presented in this paper.


\bsp	
\label{lastpage}
\end{document}